\documentstyle[12pt,psfig,epsfig,supertabular]{article}
\def\aprle{\buildrel < \over {_{\sim}}}

\begin{document}     
\title{\bf Comparison of the FLUKA calculations with CAPRICE94 data on muons in atmosphere}
\author{G.~Battistoni$^a$, A.~Ferrari$^{a,b}$, T.~Montaruli$^c$ 
\\ 
P.~R.~Sala$^{a,b}$ } 
\date{	}
\maketitle
 
\begin{center}
$a)$ INFN and Dipartimento di Fisica dell' Universit\'a, 20133 Milano, Italy \\
$b)$ now at CERN, Geneva, Switzerland \\
$c)$ INFN and Dipartimento di Fisica dell' Universit\'a, 70126 Bari, Italy
\\
and INFN Laboratori Nazionali di Frascati, 00044 Frascati, Italy
\date{}
\maketitle
 
\end{center}
 
\begin{abstract}
 
In order to benchmark the 3-dimensional calculation of the 
atmospheric neutrino flux based on the FLUKA Monte Carlo code,  
muon fluxes in the atmosphere have been computed 
and compared with data taken by
the CAPRICE94 experiment
at ground level and at different altitudes in the
atmosphere. For this purpose only two additions have been introduced 
with respect to the neutrino flux calculation: the specific solar modulation 
corresponding to the period of data taking and the bending of charged
particles in the atmosphere. Results are in good agreement with experimental
data, although improvements in the model are possible.
At this level, however, it is not possible to disentangle the interplay
between the primary flux and the interaction model.
\end{abstract}
 
\section{Introduction}
Recently, a new atmospheric neutrino flux calculation, based on the FLUKA
Monte Carlo model\cite{fluka}, has been proposed\cite{flukanu}. 
This new calculation has been produced with the aim to introduce a high
level of refinement in the modeling of all aspects of shower calculations
in atmosphere, from particle production to particle transport. The need for
a 3-Dimensional approach has also been demonstrated.
For example, considering as reference the Bartol flux
calculation\cite{bartol}, 
and accounting for the differences in the 1-D and 3-D approaches,
the FLUKA results are, for $E_\nu>$ 1 GeV, $\sim$15\% lower.
This has been obtained for the same input primary spectrum and
the reduction is present also at energies 
far from effects due to geomagnetic  
field.
This was attributed to the differences between particle production
models\cite{now2000}. 
An almost identical difference is present
with respect
to the Honda et al. simulation\cite{HKKM} (HKKM in the following), but in
this case the input primary spectrum is not the same and is believed to be
the main source of the discrepancy.

Since at present there is not yet a firm experimental determination of the
absolute neutrino flux, 
one of the possible ways to check the validity of shower calculations,
and of the corresponding interaction models, is to
compare the results for other detectable particles, like
for instance muons. Muons have a direct link with
neutrinos through the decay mechanisms, which are rather well known.
Unlike neutrinos, not all muons produced in atmospheric
showers are available at ground level, due to energy loss and decay
along their path. 
Therefore, the data taken at different altitudes by means of balloon--born
detectors are of particular interest. 
In this respect, the data obtained by the CAPRICE experiment
are of particular interest and considered among the most reliable
 by the scientific community.
We are interested in particular to the data taken at
Lynn Lake in 1994 and published in \cite{caprice94}.
In this paper we describe the results of the comparison of our FLUKA
calculations with this data set, showing that, under the assumption
concerning the validity of the input primary spectrum adopted so far, 
a good agreement with these muon data is achieved.

Another neutrino flux calculation has been recently proposed\cite{naumov},
with a model capable of obtaining a successful comparison with the CAPRICE
data. In this case the normalization of the resulting neutrino flux
is different from the FLUKA one, depending on neutrino flavor and energy,
but with the general trend of producing a 20 -- 25\% lower normalization 
in the sub--GeV region. According to the authors, when switching from their
particle production model to the Bartol one the resulting neutrino fluxes
obtained with their code (CORT) increase by a factor exceeding 50\%
in the Sub--GeV range.
They conclude that their model, which is able to reproduce 
the muon fluxes in the atmosphere, predicts significantly lower neutrino
fluxes than those currently used in the experimental analyses
(mostly the Bartol\cite{bartol} and HKKM\cite{HKKM} computations).

\section{Monte Carlo setup}
The FLUKA code \cite{fluka} is a general purpose Monte Carlo code
for the interaction and transport of particles.
It is built and maintained with the aim of including the best possible 
physical models in terms of completeness and precision.
It contains detailed
models of electromagnetic, hadron--hadron and hadron--nucleus interactions, 
covering the range from the MeV scale to the many--TeV one. 
Extensive benchmarking against experimental data has been produced (see the 
references in  \cite{fluka}).
In view of applications for cosmic ray physics, we have implemented in the
FLUKA environment a 3--Dimensional spherical representation of the whole
Earth  and of the surrounding atmosphere. 
The latter is described by a 
by a proper mixture of N, O and Ar, arranged in 100 concentric
spherical shells, whose densities follow
the known profile of the USA ``standard atmosphere".
This is the standard choice in neutrino flux calculations,
although we are aware
that important differences exist as a function of latitude and season.
For a muon flux calculation the specific
features of the local geographical site and the actual situation on the
days of measurement could be relevant for a fully significant comparison
with data. We have also tried to set the boundaries between the atmospheric
shells in such 
a way to reproduce as much as possible the different ``flux weighted
average depths'' at which the
CAPRICE94 measurements were reported. Considering our discrete structure
and the adopted atmosphere profile, we evaluated a systematic error
on the atmospheric depth of less than few percent.

\begin{figure}[thb]
\begin{center}
\mbox{\epsfig{file=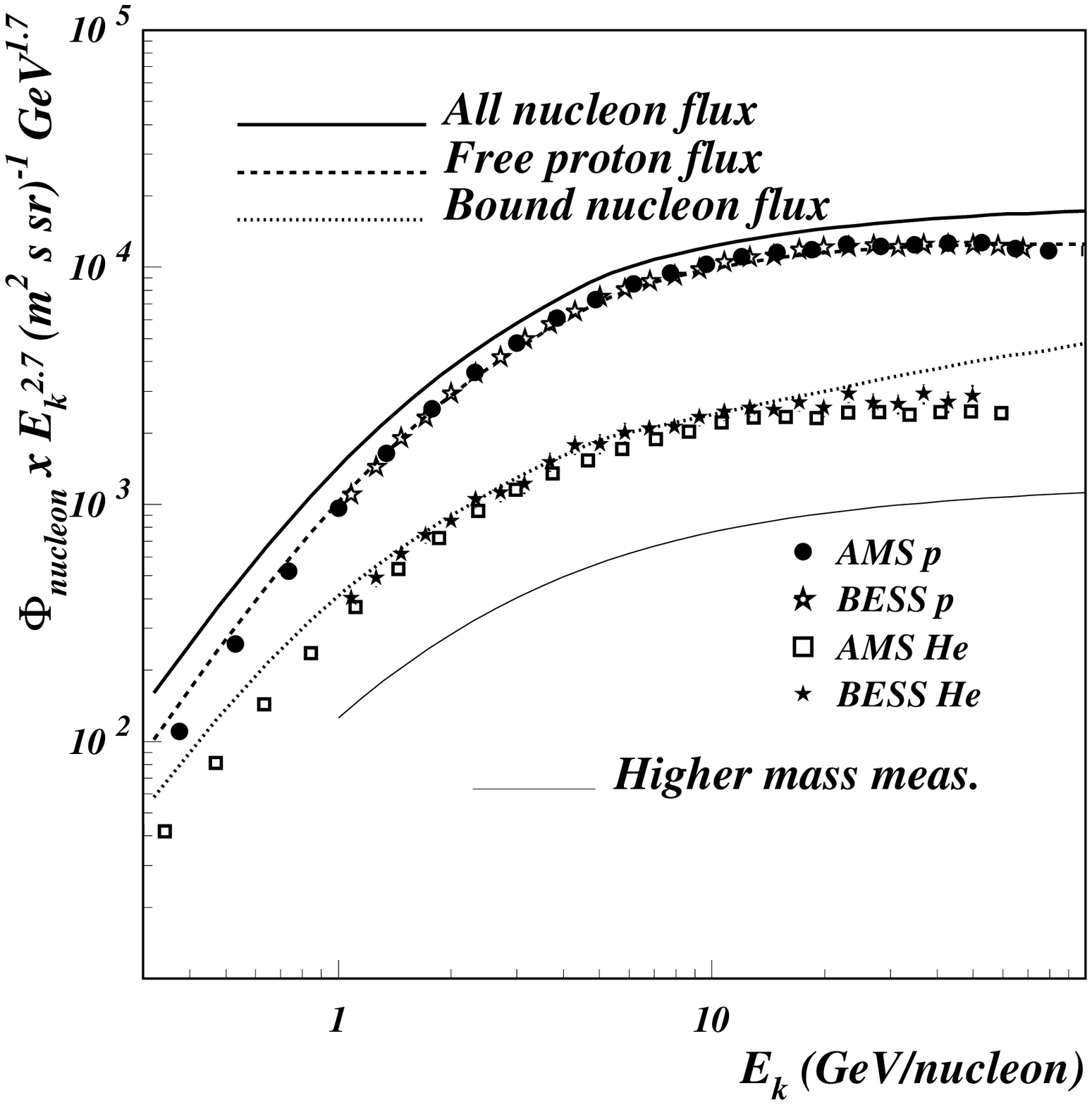,width=12cm}} 
\caption{All-nucleon primary spectrum (from Bartol group) used for the FLUKA
calculations in the atmosphere as compared to recent AMS and BESS data
sets. The continuous line representing higher mass components has been
derived from ref. \protect\cite{wiebel}.
\label{fig:prim}} 
\end{center}
\end{figure}

The starting point of any calculation of particle fluxes in the atmosphere is 
the primary flux. In order to allow a direct comparison with previously
published computations, we used the fit of
all--nucleon flux proposed by the Bartol group\cite{bartol}.
This flux has also the merit that, at least up to about hundred GeV 
it is quite close to the recent measurements of proton primaries
performed by BESS\cite{bess} and AMS\cite{ams}. A comparison with 
these data sets is shown in Fig. \ref{fig:prim}, where the all--nucleon
spectrum has been divided in the free--proton and bounded--nucleon
components. To the bounded--nucleon component, mass groups heavier than
helium contribute. The sum of such contributions, taken from 
ref. \cite{wiebel}, are represented by a continuous line.
Nuclei are treated in the framework of the super-position model, namely a
nucleus of mass number A, and total energy E$_0$ is considered to be
equivalent to A nucleons each having energy E$_0$/A.
At present, this is the approach followed in all other neutrino
calculations, except the case of ref.\cite{naumov}, where a Glauber-like
model is used for the primary nuclear component.

\begin{figure}[thb]
\begin{center}
\mbox{\epsfig{file=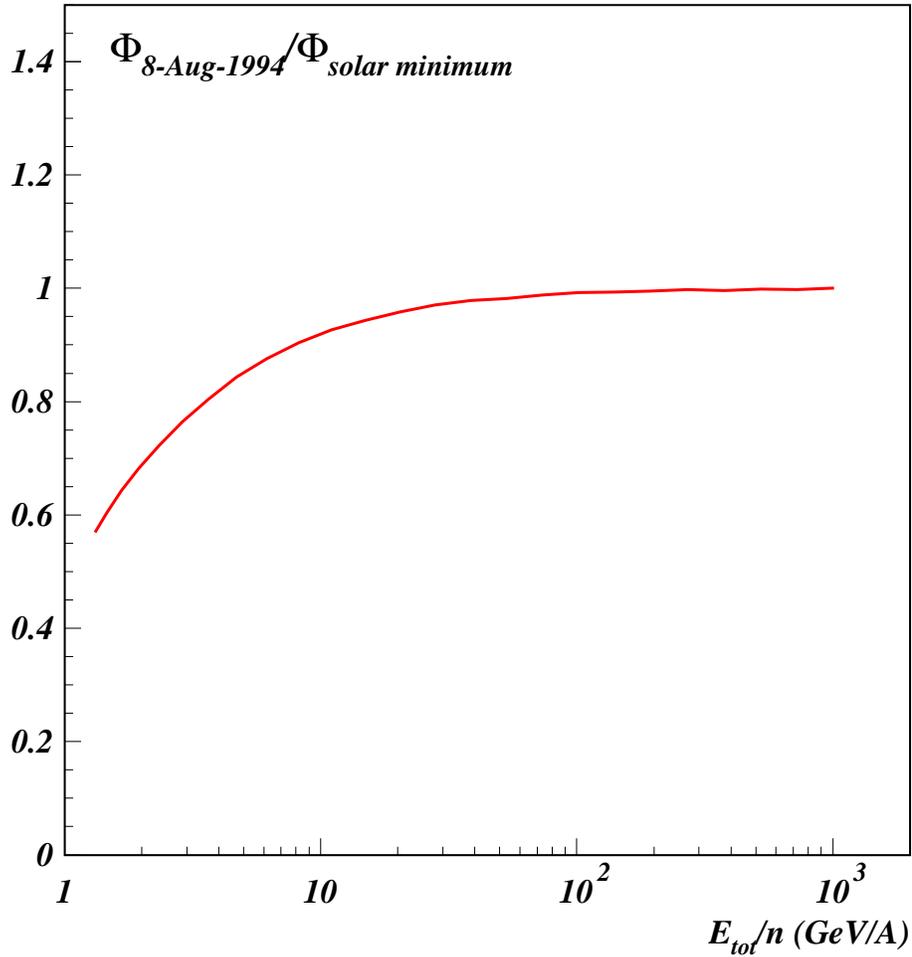,width=12cm}} 
\caption{Ratio of all-nucleon primary spectrum modulated for the date of
August 8$^{th}$ 1994 to that at solar minimum value, as obtained using the
algorithm of ref. \protect\cite{modulation}.
\label{fig:modsol}} 
\end{center}
\end{figure}

For the purpose of the present calculation, we have modified the spectrum
in order to take into account the solar modulation at the time of the
experimental measurement (August 8th 1994). In order to do that, we have
followed the modeling of ref. \cite{modulation}. First, a
Local Interstellar Spectrum has been derived from the Bartol spectrum
at solar minimum, and than it has been modulated back to the desired level
according to the counting rate of the Climax neutron monitor.
The effect of solar modulation on all--nucleon spectrum is shown
in Fig. \ref{fig:modsol}.
Primary particles, sampled
from the continuous energy spectrum, are injected
at the top of the atmosphere, 
at about 100 km of altitude. 

The primary flux is assumed to be 
uniform and isotropic far from the Earth, and it is locally modified by
the effects of the geomagnetic
field\cite{geomag} and solar modulation. 
Contrary  to our 
previous neutrino flux 
calculation we have switched on the local magnetic field also during
shower development, in order to allow the bending of charged particle
trajectories.

In the simulation we score the muon flux separately for each muon
charge, at each boundary crossing 
between the different atmospheric shells, and
at the Earth boundary as well. 
We have tested different acceptance cones to
define the fluxes, using half-cone angles of 8, 12 and 20 degrees. At a few
altitudes we have also added a cone of 50 degrees of half-aperture.
The muon flux is recorded as function of laboratory momentum, in 30 logarithmic
bins between 0.2 and 200 GeV/c. In order to compare our results to the
experimental data, a binning of calculated fluxes has been performed
to reproduce the momentum ranges considered in \cite{caprice94}

We have also performed simulation runs in the 1--Dimensional approximation,
that is aligning all secondary particles to the primary at each interaction.

\section{Results}

\begin{figure}[thb]
\begin{center}
\begin{tabular}{c}
\mbox{\epsfig{file=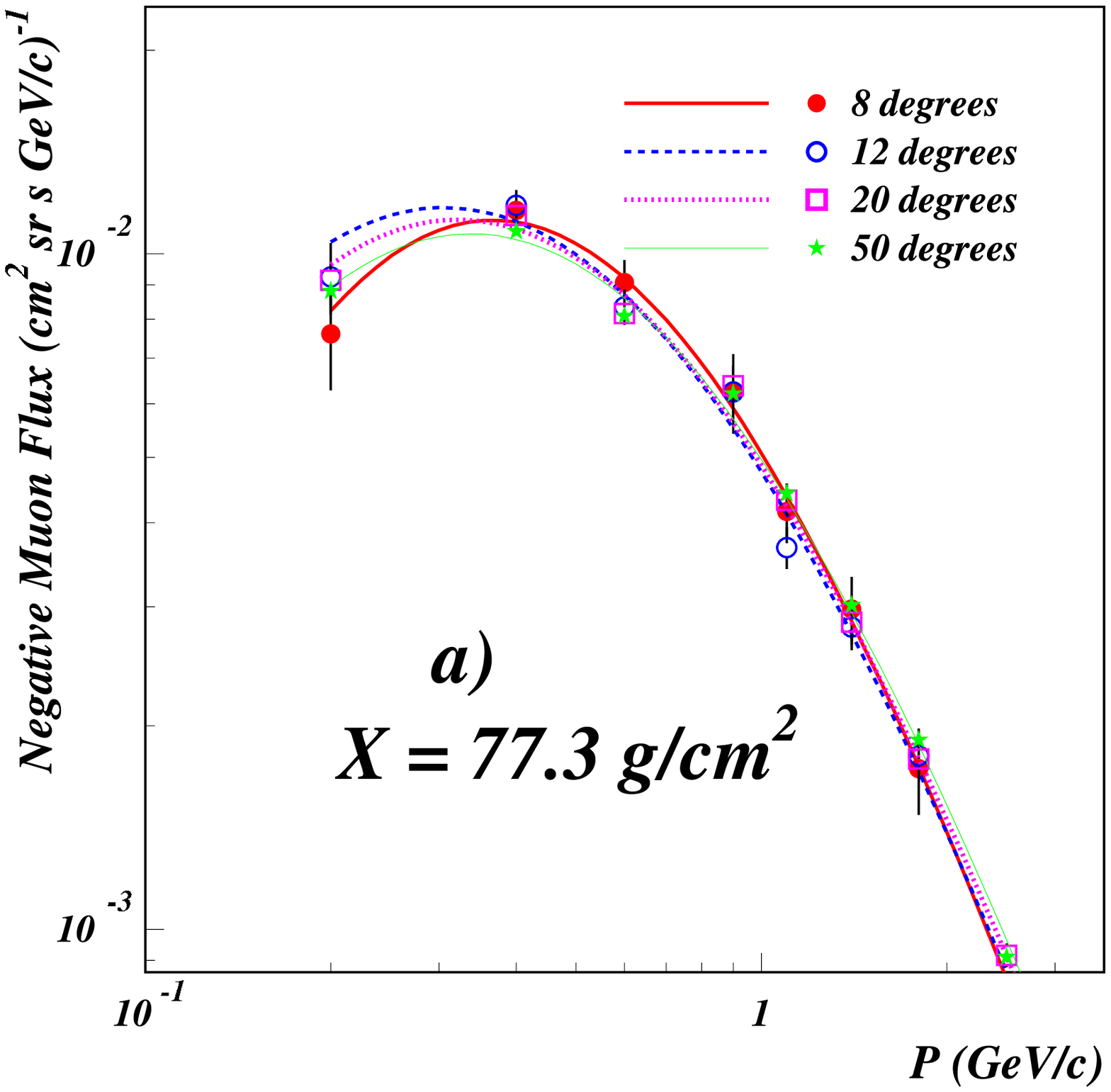,width=9cm}} \\
\mbox{\epsfig{file=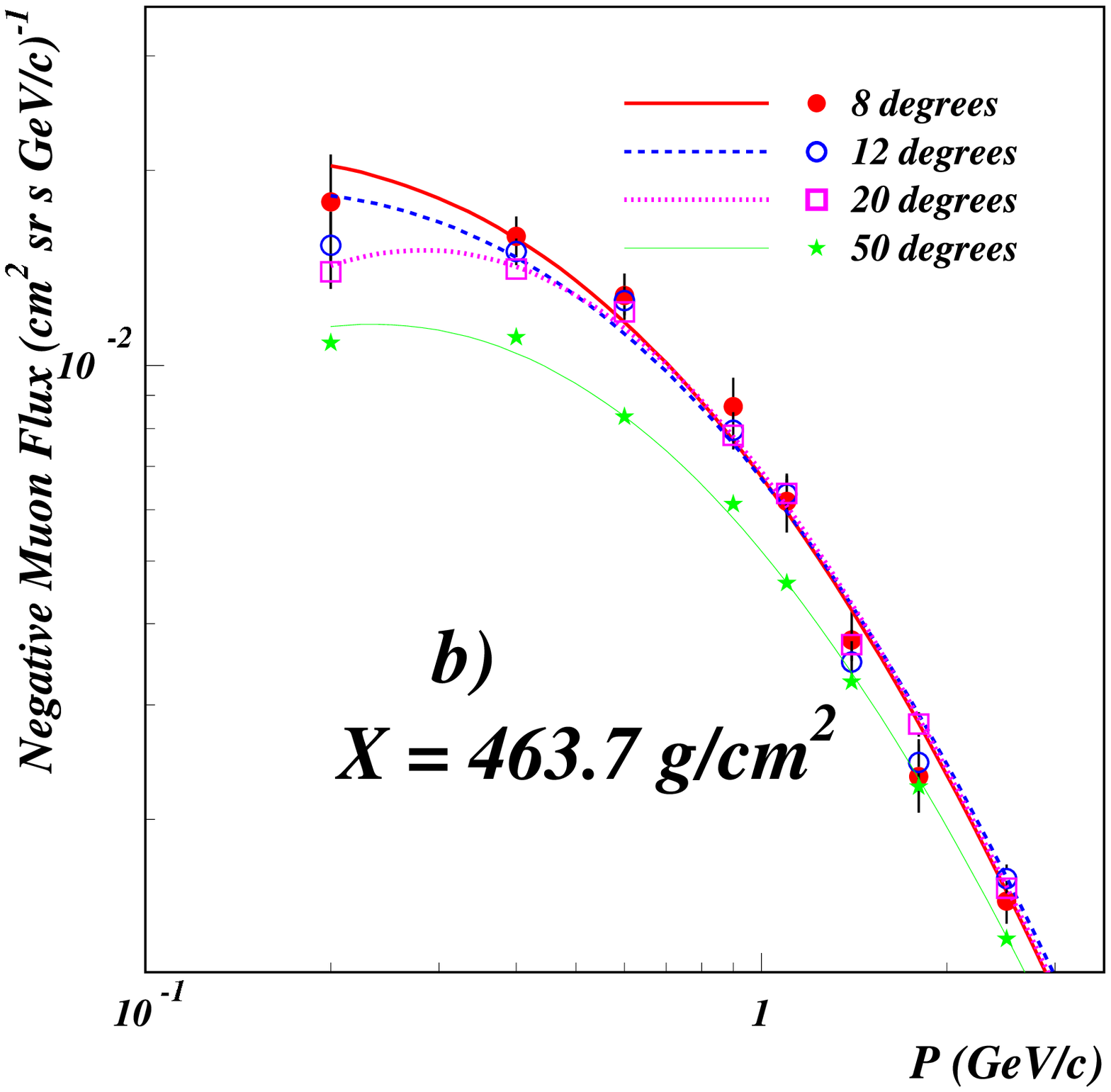,width=9cm}} 
\end{tabular}
\caption{Calculated Negative muon flux
as a function of momentum for different aperture cones is shown at the
atmospheric depth of 50.7 g/cm$^2$ (a), corresponding to about 20.6 km of
altitude (a) 
and at the depth of 308.9 g/cm$^2$ (b), that is at about 9.08 km of 
altitude). The lines are phenomenological fits meant to guide
the eye.\label{fig:angdist}} 
\end{center}
\end{figure}

As a first step, we have studied the dependence of simulated fluxes, at
different heights, as a function of the different aperture cones.
As a result we may conclude that up to 20 degrees, the simulation does not 
show a significant dependence on the aperture cone at all heights.
In the 50 degrees cone, instead, a loss of muon flux at low muon momentum
can be observed with respect to the
lower apertures. As an example, in Fig. \ref{fig:angdist} the negative muon
flux as a function of momentum for different aperture cones is shown at the
nominal 
atmospheric depth of 50.7 g/cm$^2$ (a), corresponding to about 20.6 km of
altitude (a) 
and at the nominal depth of 308.9 g/cm$^2$ (b), that is at about 9.08 km of
altitude.
The loss at low muon momentum at large angle is explained by muon energy
loss and decay.

\begin{figure}[htb]
\begin{center}
\begin{tabular}{cc}
\hspace{-1cm}
\mbox{\epsfig{file=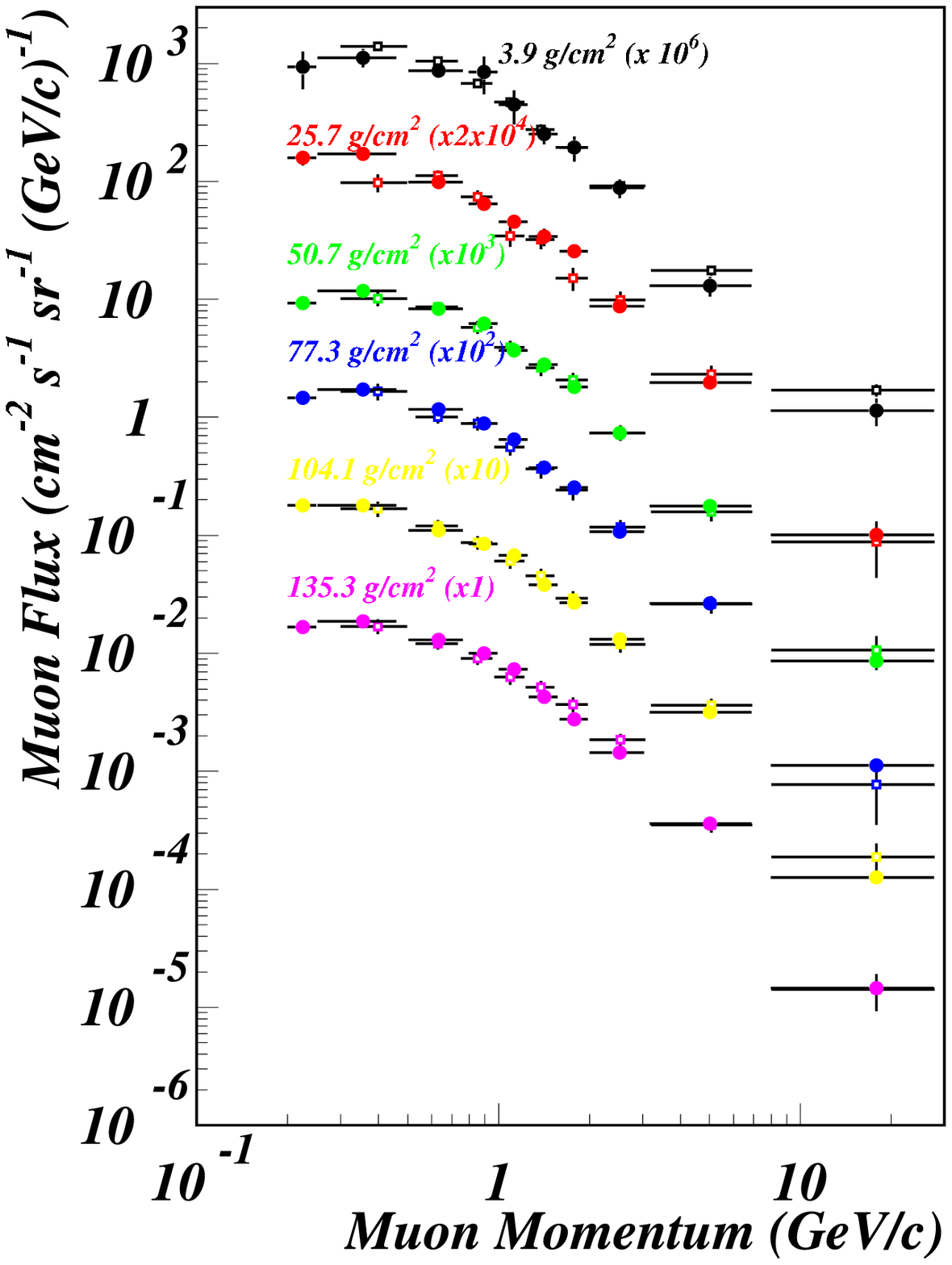,height=12cm}} &
\mbox{\epsfig{file=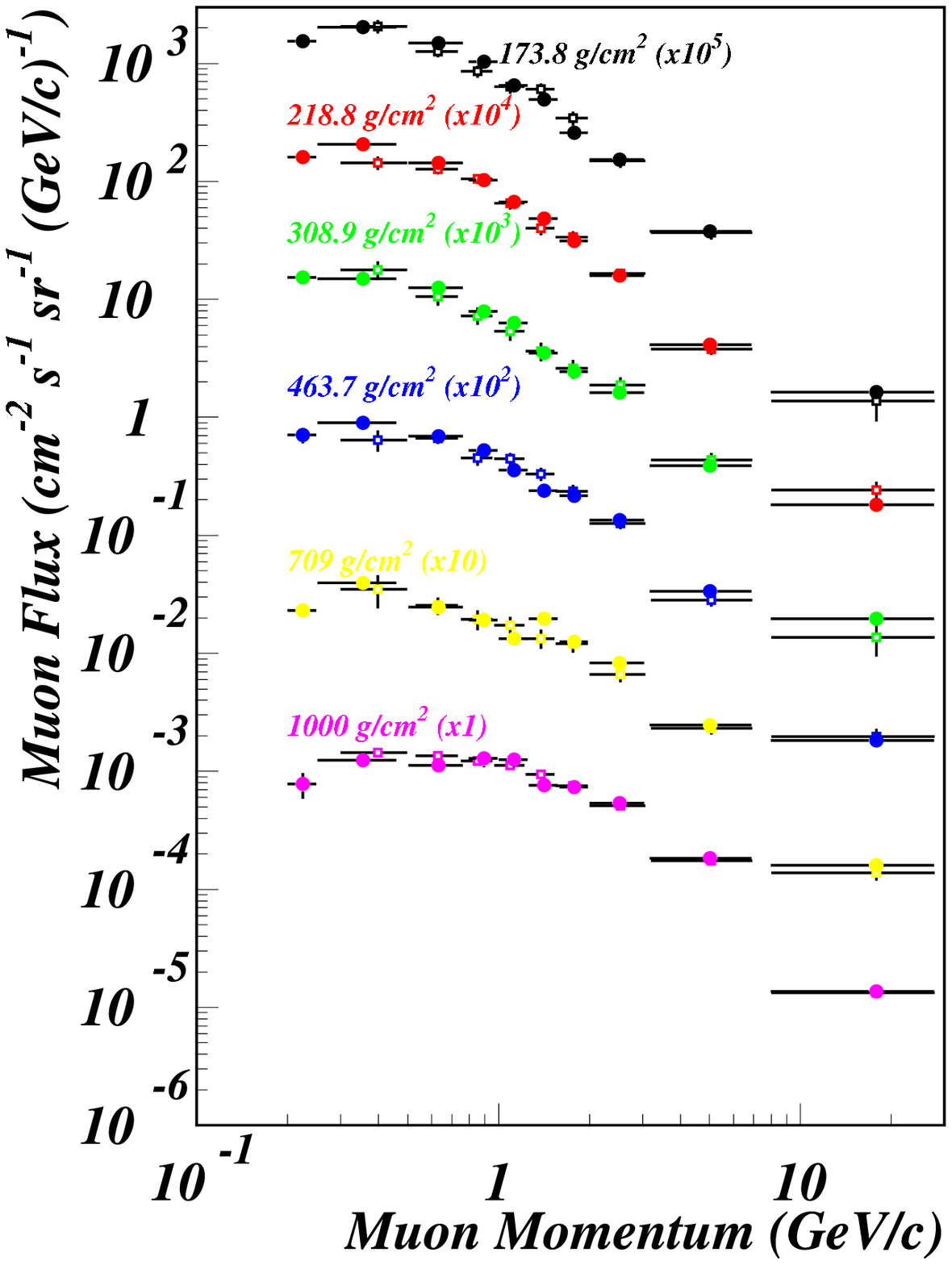,height=12cm}} 
\end{tabular}
\caption{Comparison between simulated and detected negative muon flux
as a function of momentum for different 
atmospheric depths.\label{fig:negm}}
\end{center}
\end{figure}

\begin{figure}[htb]
\begin{center}
\begin{tabular}{cc}
\hspace{-1cm}
\mbox{\epsfig{file=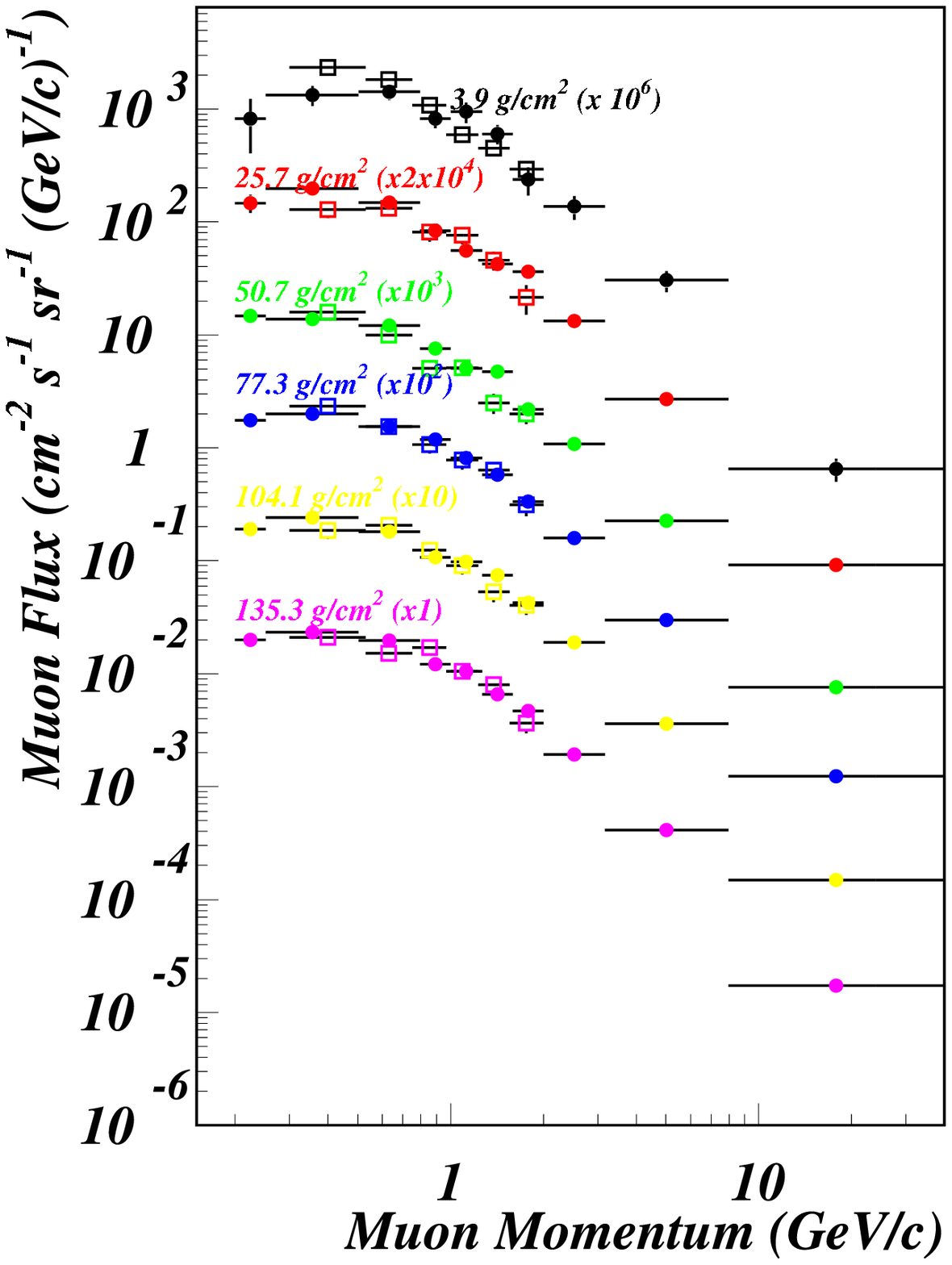,height=12cm}} &
\mbox{\epsfig{file=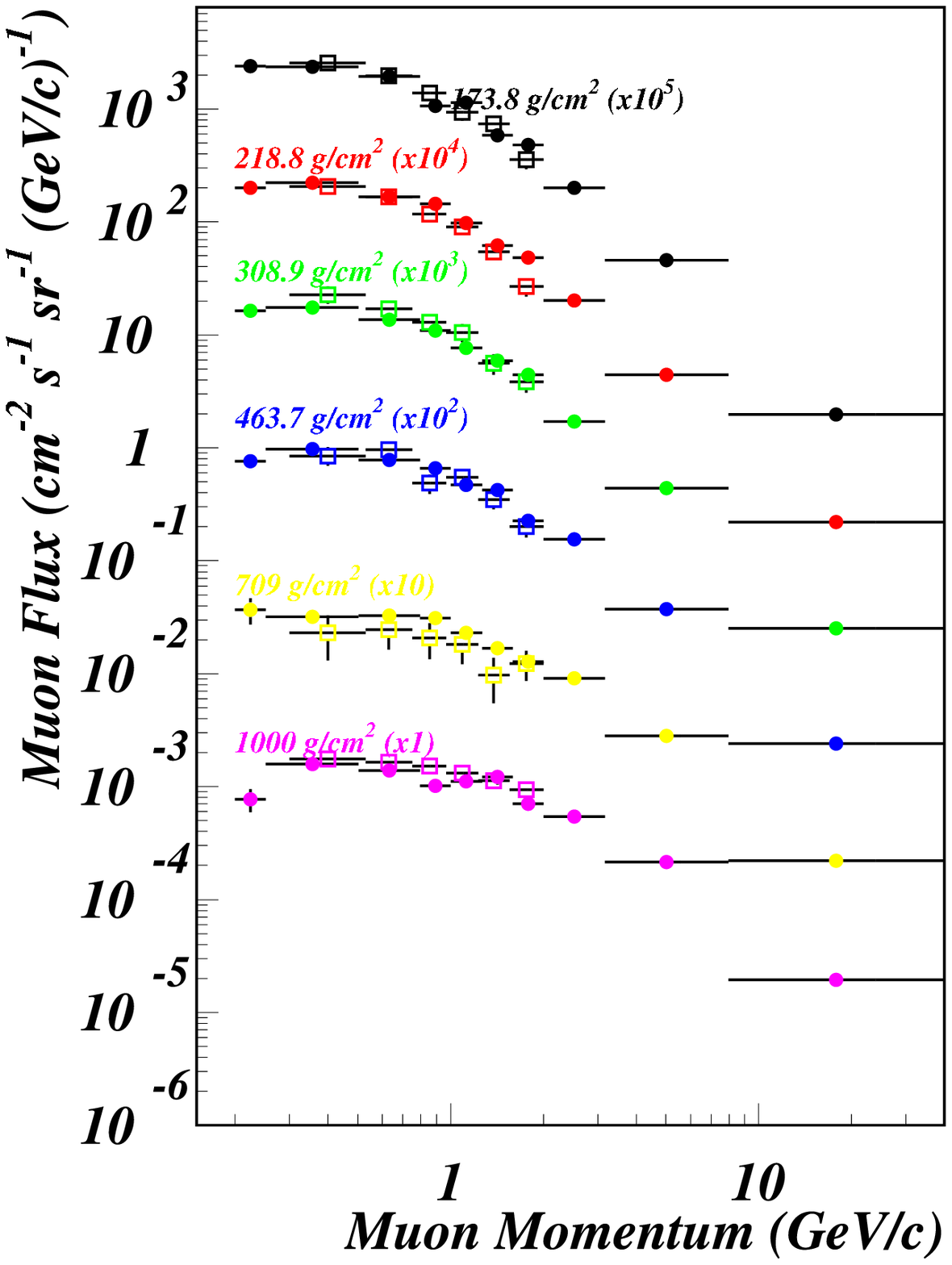,height=12cm}} 
\end{tabular}
\caption{Comparison between simulated (small full symbols) and detected
positive muon flux (open symbols)
as a function of momentum for different 
atmospheric depths.\label{fig:posm}}
\end{center}
\end{figure}

\begin{figure}[htb]
\begin{center}
\hspace{-1cm}
\mbox{\epsfig{file=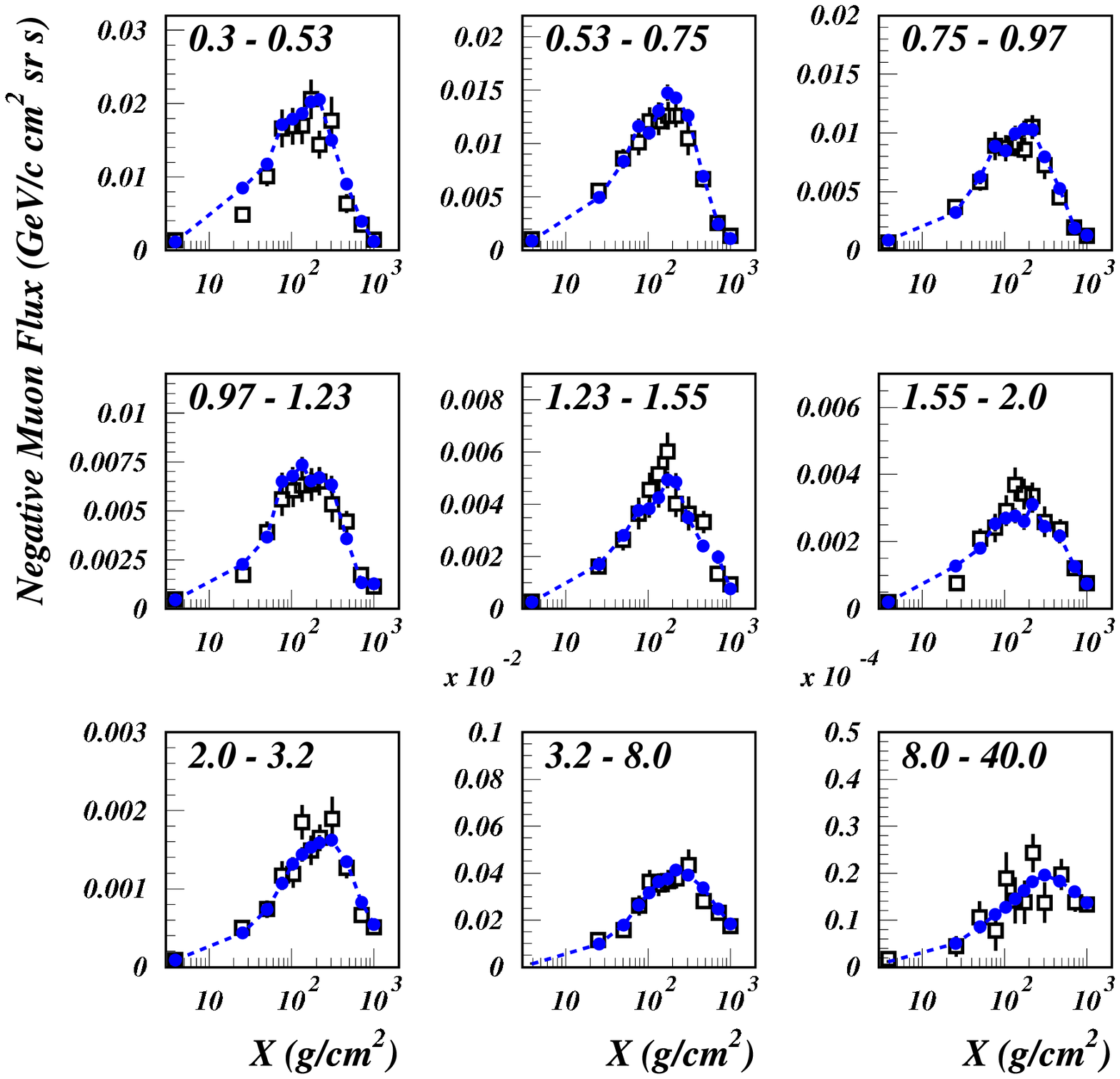,height=12cm}}
\caption{Comparison between simulated (full symbols) and detected
negative muon flux (open symbols)
as a function of atmospheric depth, for different 
momentum bins. The dotted line is draw to guide the eye. \label{fig:myneg}}
\end{center}
\end{figure}

\begin{figure}[htb]
\begin{center}
\hspace{-1cm}
\mbox{\epsfig{file=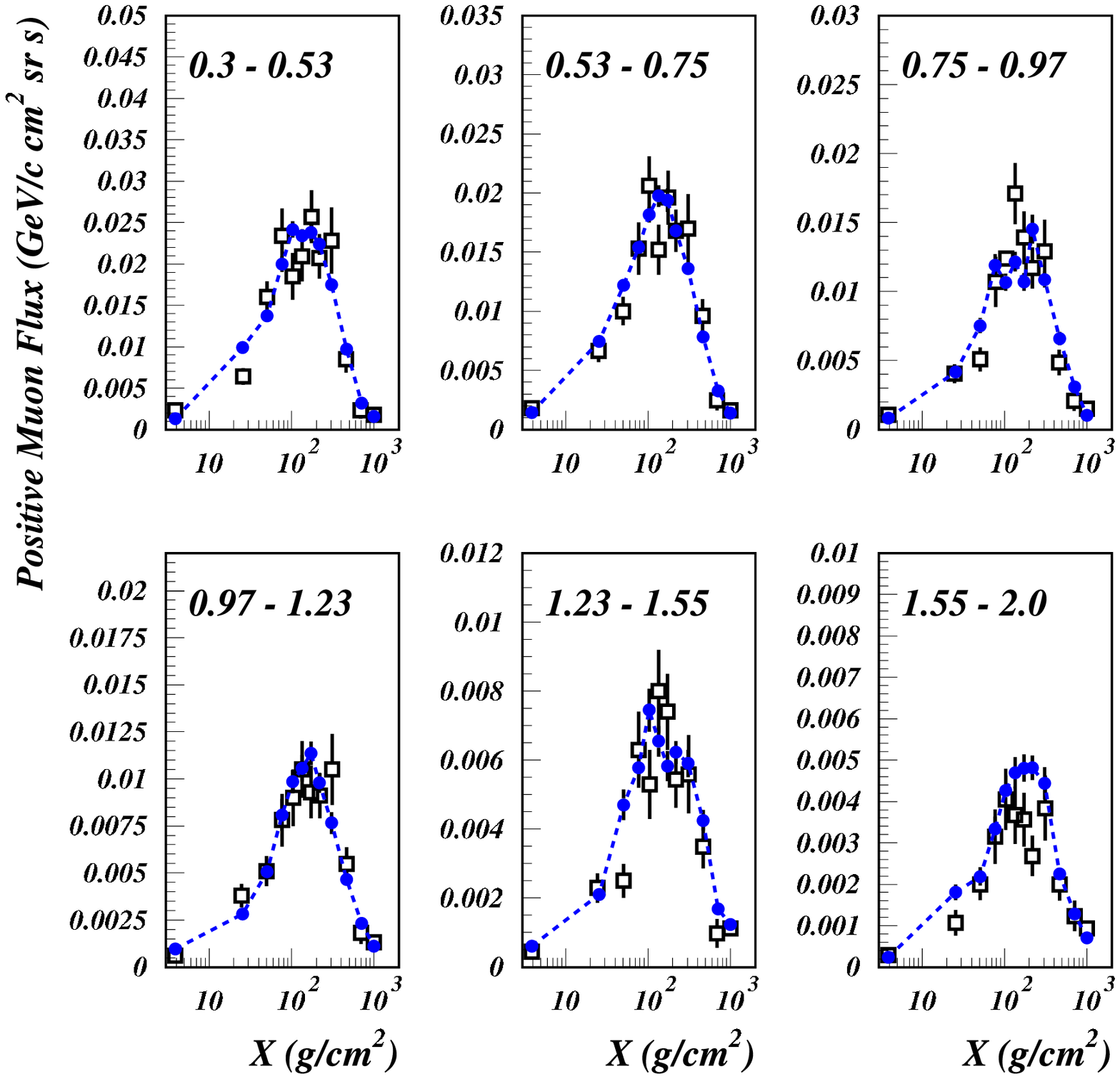,height=12cm}} 
\caption{Comparison between simulated (full symbols) and detected
positive muon flux (open symbols)
as a function of atmospheric depth, for different 
momentum bins. The dotted line is drawn to guide the eye.\label{fig:mypos}}
\end{center}
\end{figure}

The comparison with the experimental data from CAPRICE is performed
using the muon flux recorded in the 12 degrees half-aperture cone, the one
closer to the apparatus acceptance.
The results, relative to the full 3--Dimensional simulation runs, are
summarized in Fig. \ref{fig:negm} for the negative muon 
measurements at different depths in atmosphere and in Fig. \ref{fig:posm}
for the positive muons. The numeric values of the simulated fluxes
reported in these figures are reported in Tables \ref{tab1} and \ref{tab2}
in the Appendix, for the different detector depths.
The same data are reported as a function of depth, in different momentum
intervals, in Fig. \ref{fig:myneg} and \ref{fig:mypos}, for negative and
positive muons respectively. 
In order to express numerically the
level of agreement between data and simulation, we report 
in table \ref{tabdep} the weighted
average of ratios of Monte Carlo to experimental data at each atmospheric
depth. as from Fig. \ref{fig:negm} and \ref{fig:posm}, (thus integrating on 
all momentum bins). In a similar way, table \ref{tabgr} shows
the average ratios at each
momentum bin (integrating on all atmospheric depths). Statistical
errors have been used to calculate the weights. 

\begin{table}[p]
\begin{center}
\begin{tabular}{|c|c|c|}
\hline
Atmospheric Depth (g/cm$^2$) & MC/Exp. Data ($\mu^-$) & MC/Exp. Data ($\mu^+$) \\ 
\hline
3.99  & 0.84  $\pm$ 0.06  & 	0.78 $\pm$ 0.07 \\
25.7  & 0.97  $\pm$ 0.07  & 	1.00 $\pm$ 0.08 \\
50.7  & 1.00  $\pm$ 0.05  & 	1.06 $\pm$ 0.07 \\
77.3  & 1.04  $\pm$ 0.06  & 	0.96 $\pm$ 0.07 \\
104.1 & 0.94  $\pm$ 0.05  & 	0.91 $\pm$ 0.04 \\
135.3 & 0.94  $\pm$ 0.05  & 	0.92 $\pm$ 0.06 \\
173.8 & 0.98  $\pm$ 0.05  & 	0.92 $\pm$ 0.06 \\
218.8 & 1.03  $\pm$ 0.04  & 	1.11 $\pm$ 0.07 \\
308.9 & 0.97  $\pm$ 0.06  & 	0.83 $\pm$ 0.07 \\
463.7 & 0.96  $\pm$ 0.05  & 	0.97 $\pm$ 0.08 \\
709   & 1.04  $\pm$ 0.06  & 	1.27 $\pm$ 0.19 \\
1000  & 0.99  $\pm$ 0.03  & 	0.79 $\pm$ 0.04 \\
\hline
\end{tabular}
\caption{Weighted average ratios of simulated
to experimental data for different atmospheric depths
and integrating on the different momentum ranges.
\label{tabdep}}
\end{center}
\end{table}

\begin{table}[p]
\begin{center}
\begin{tabular}{|c|c|c|}
\hline
Muon momentum (GeV/c) & MC/Exp. Data ($\mu^-$) & MC/Exp. Data ($\mu^+$) \\ 
\hline
 0.3--0.53  & 1.02 $\pm$ 0.05 &  0.92 $\pm$ 0.04 \\ 
0.53--0.75  & 0.97 $\pm$ 0.03 &  0.94 $\pm$ 0.04 \\
0.75--0.97  & 1.04 $\pm$ 0.05 &  0.83 $\pm$ 0.03 \\
0.97--1.23  & 1.00 $\pm$ 0.05 &  0.94 $\pm$ 0.05 \\
1.23--1.55  & 0.88 $\pm$ 0.04 &  1.00 $\pm$ 0.06 \\
1.55--2.0   & 0.91 $\pm$ 0.04 &  0.98 $\pm$ 0.06 \\
2.0--3.2    & 0.97 $\pm$ 0.04 &	----	 \\
3.2--8.0    & 1.02 $\pm$ 0.03 &	----	 \\
8.0--40.0   & 0.96 $\pm$ 0.05 &	----	 \\
\hline
\end{tabular}
\caption{Weighted average ratios of simulated
to experimental data in different momentum bins
and integrated on the different atmospheric depths.
\label{tabgr}}
\end{center}
\end{table}

We consider satisfactory the general level of
agreement, also taking into account that here we are assuming that 
no systematic error exist for this set of experimental data. Instead, as
stated before, a few percent systematics in the calculation should
come from the atmospheric description.
We notice a deficit in the simulation with respect to the data
at very low momenta at the topmost altitude.
In the case of negative muons there could be at most
a 10\% deficit in simulation.
in the range 1.23$\div$2.0 GeV/c. Here the deficit
is dominated by low values obtained at the shower
depths between 100 and 170 g/cm$^2$, although both at 77 and 218 g/cm$^2$
(still close to the shower maximum) a ratio close to 1.0 is again achieved. 
For positive muons the maximum deficit is
limited to 0.75$\div$0.97 momentum bin, again dominated by the results
in the range between 100 and 170 g/cm$^2$. However,
we want to stress that, for a given
interaction model and calculation scheme, these values depend
also on the choice of primary all nucleon spectrum. 
Furthermore we also expect
that our assumption on the total absence of systematics in the 
experimental data is probably too optimistic.
With the present level of statistical error we cannot 
identify systematic problems in the $\mu^+/\mu^-$ ratio at these 
altitudes.

\begin{figure}[htb]
\begin{center}
\begin{tabular}{c}
\mbox{\epsfig{file=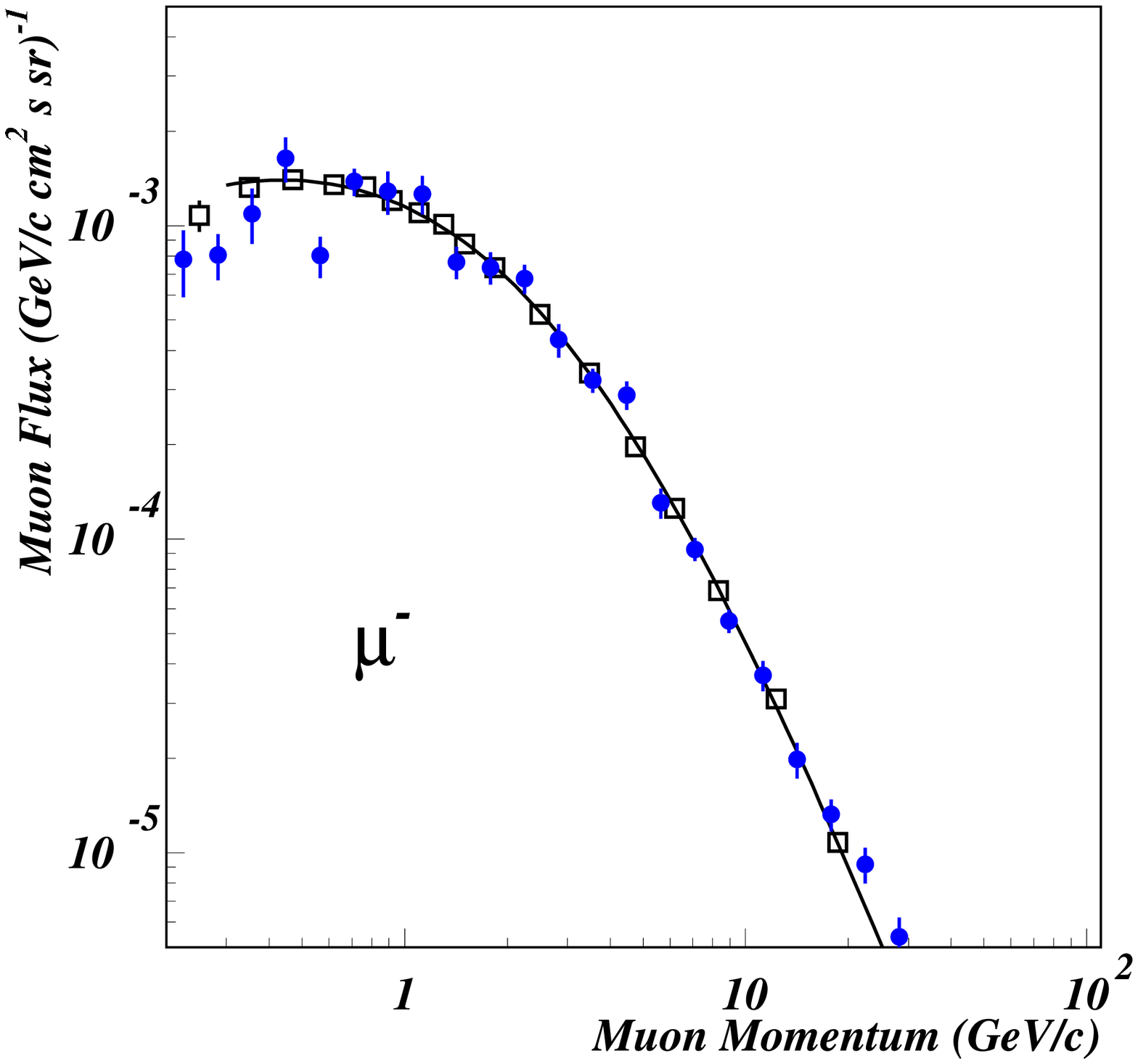,width=7cm}} \\
\mbox{\epsfig{file=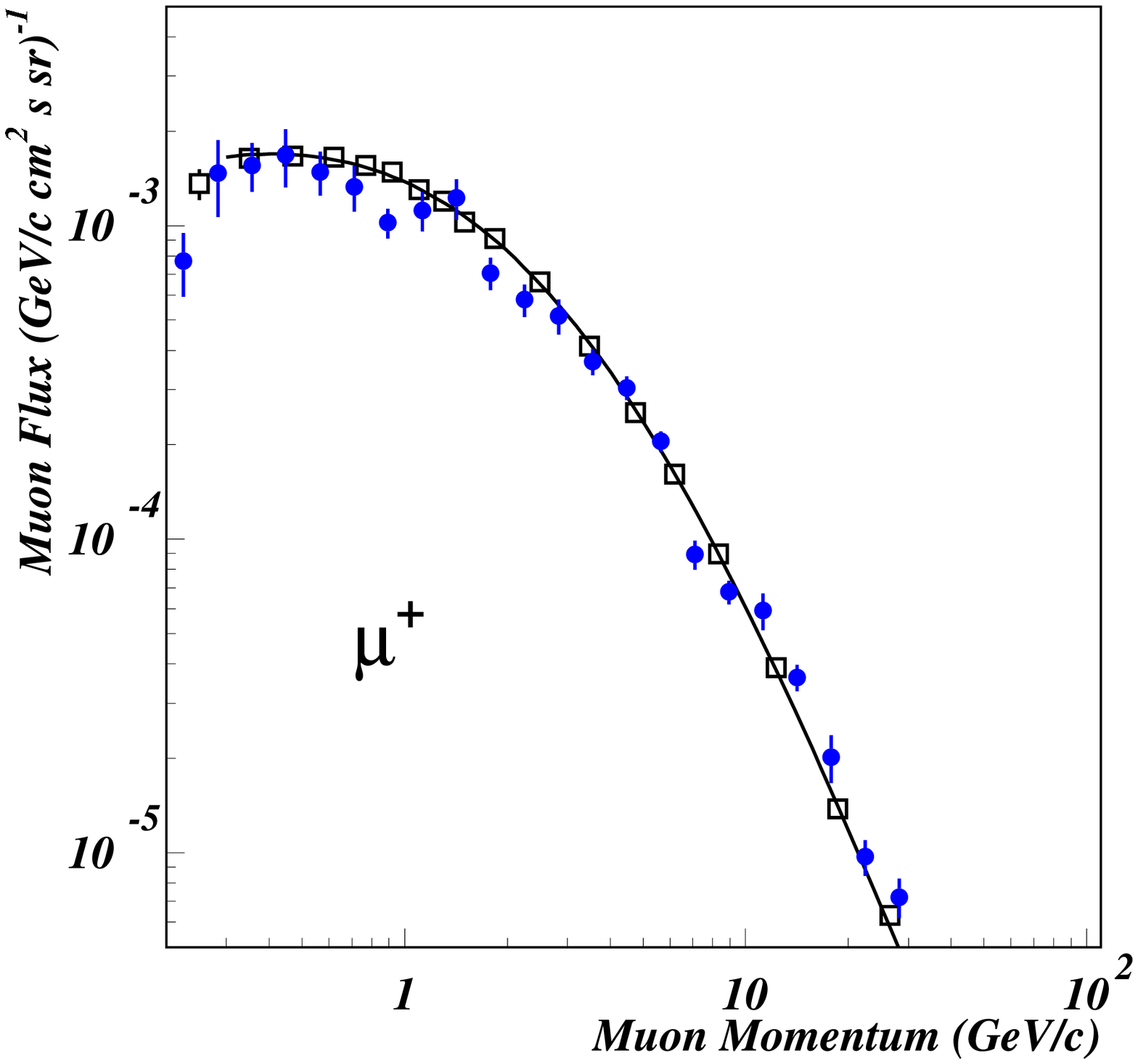,width=7cm}} 
\end{tabular}
\caption{Comparison between simulated (full symbols) and detected (open
symbols) negative and positive muon fluxes
as a function of momentum at ground level. The continuous lines
are phenomenological fits to the experimental data (above 0.3 GeV/c).
\label{fig:grnd}}
\end{center}
\end{figure}

\begin{figure}[htb]
\begin{center}
\mbox{\epsfig{file=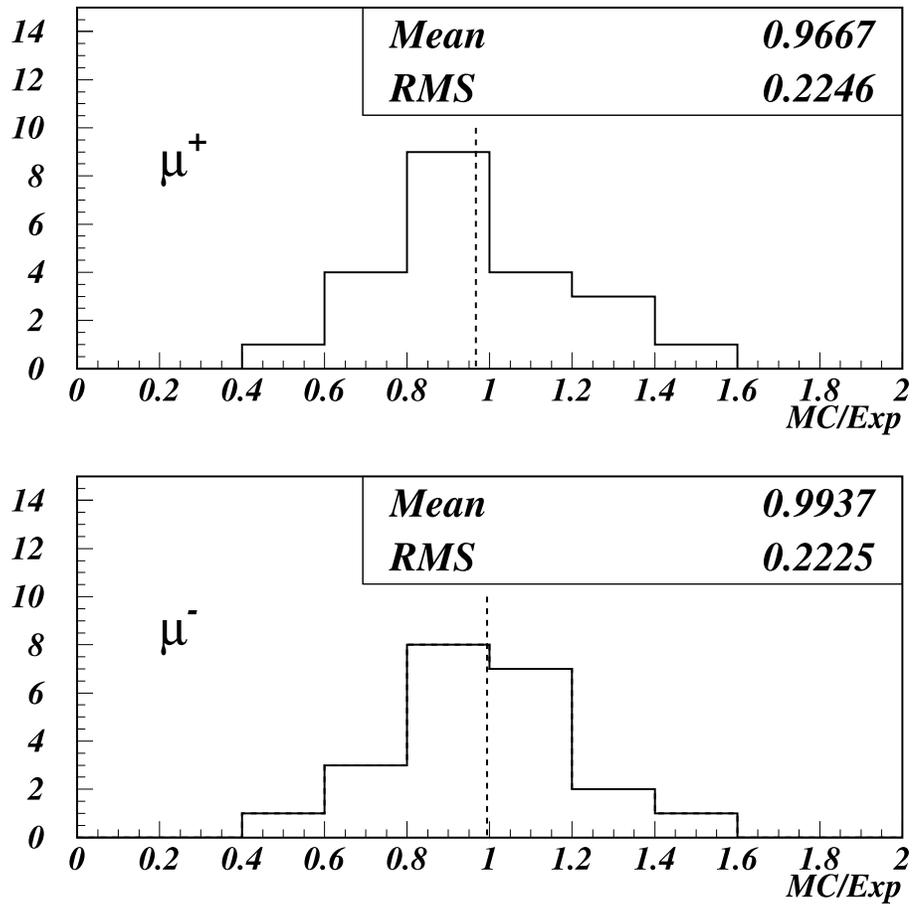,width=12cm}} 
\caption{Distribution of the ratios of calculated to experimental muon flux
at ground level for the different muon charges.
\label{fig:ratgr}}
\end{center}
\end{figure}

Fig. \ref{fig:grnd} shows the comparison with the measurement at
ground level. Numerical values are given in table \ref{tab3}. 
In this case the flux have been calculated in momentum bins which are
slightly different with respect to those of experimental data.
Therefore, in order to evaluate the level of agreement, we calculated
the ratios of simulated fluxes to the values at the same momentum (above
0.3 GeV/c) taken from the phenomenological fits to experimental data
shown as continuous lines in Fig. \ref{fig:grnd}. The distribution of
these ratios are shown in Fig. \ref{fig:ratgr}. Here, the present results
could point out a possible deficit in the calculated $\mu^+$'s with respect
to the $\mu^-$. 
However, it is again important to remember that also the 
modeling of primary spectrum does affect this result, through the 
neutron to proton ratio which in turn is determined by the relative
abundances of nuclei and free nucleons.

\begin{figure}[p]
\begin{center}
\begin{tabular}{cc}
\mbox{\epsfig{file=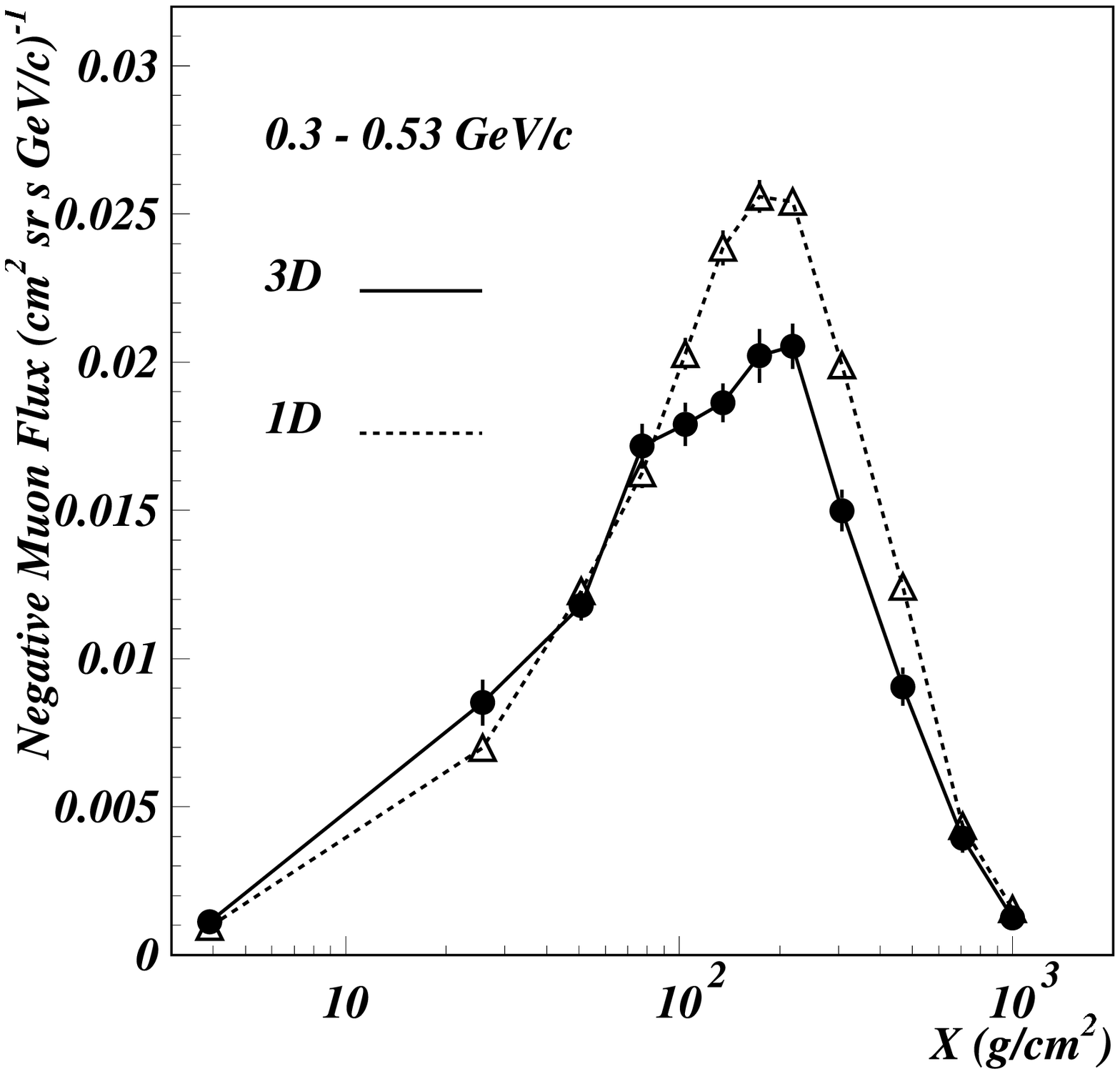,width=7cm}} &
\mbox{\epsfig{file=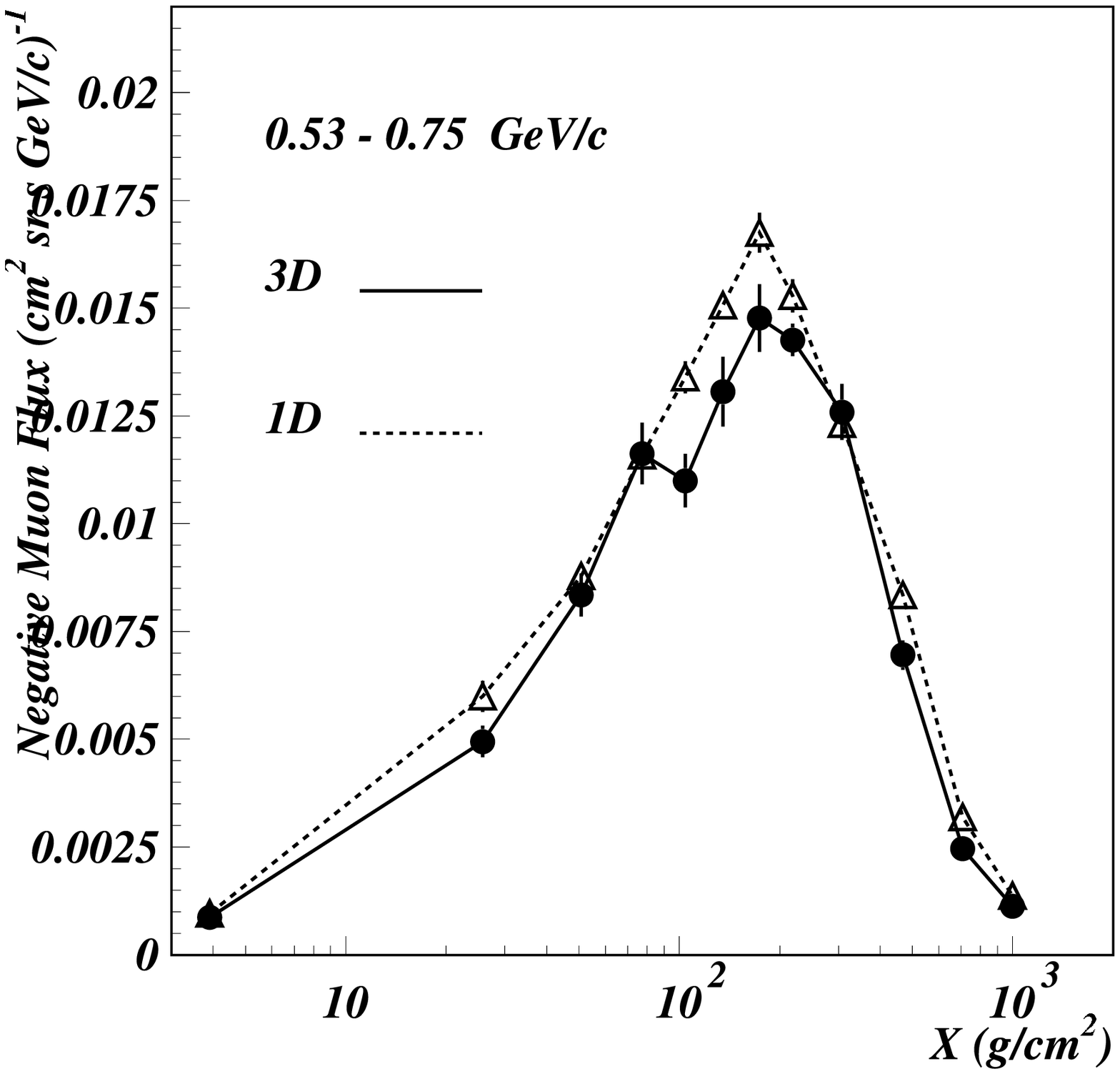,width=7cm}} \\
\mbox{\epsfig{file=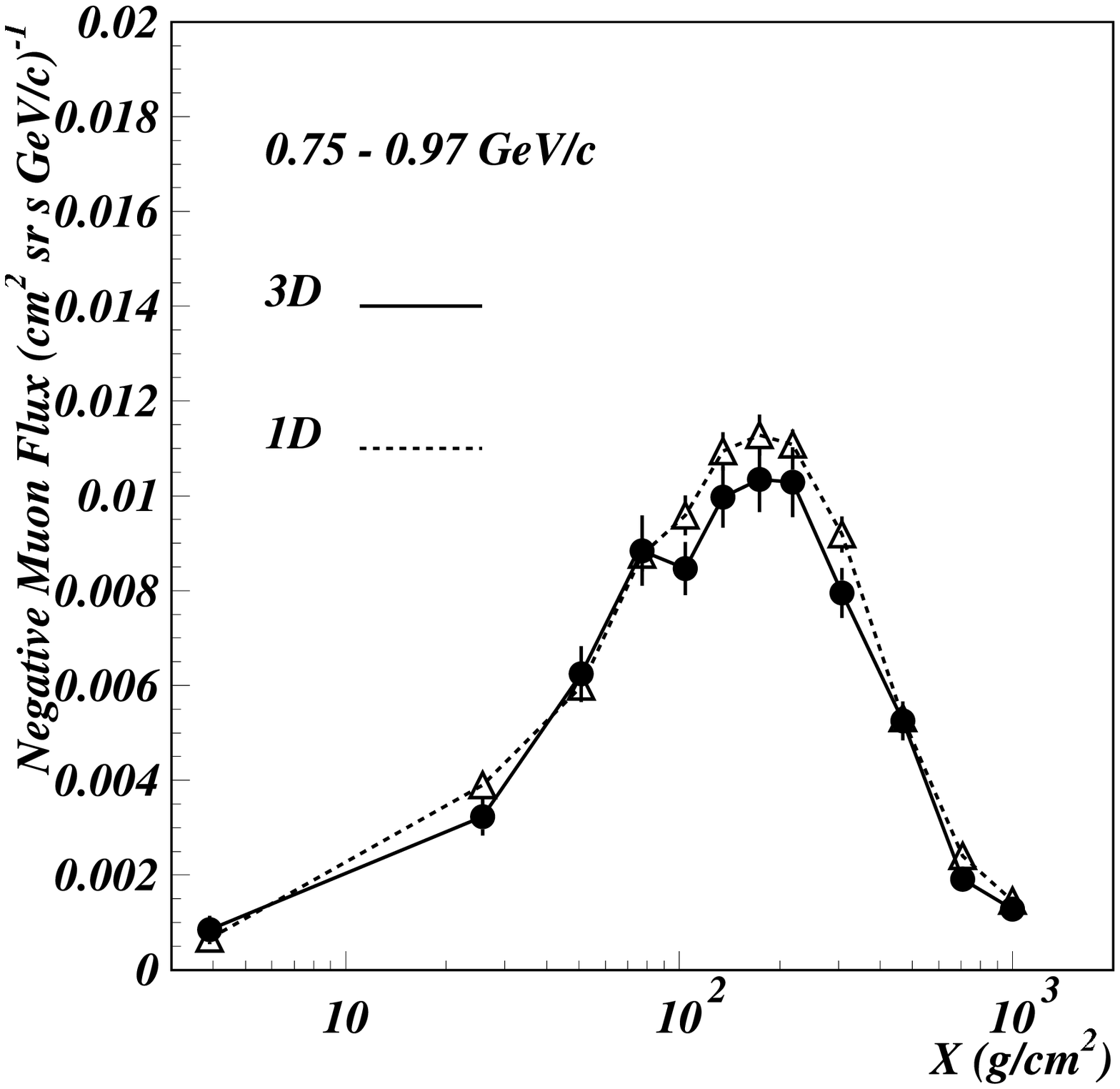,width=7cm}} &
\mbox{\epsfig{file=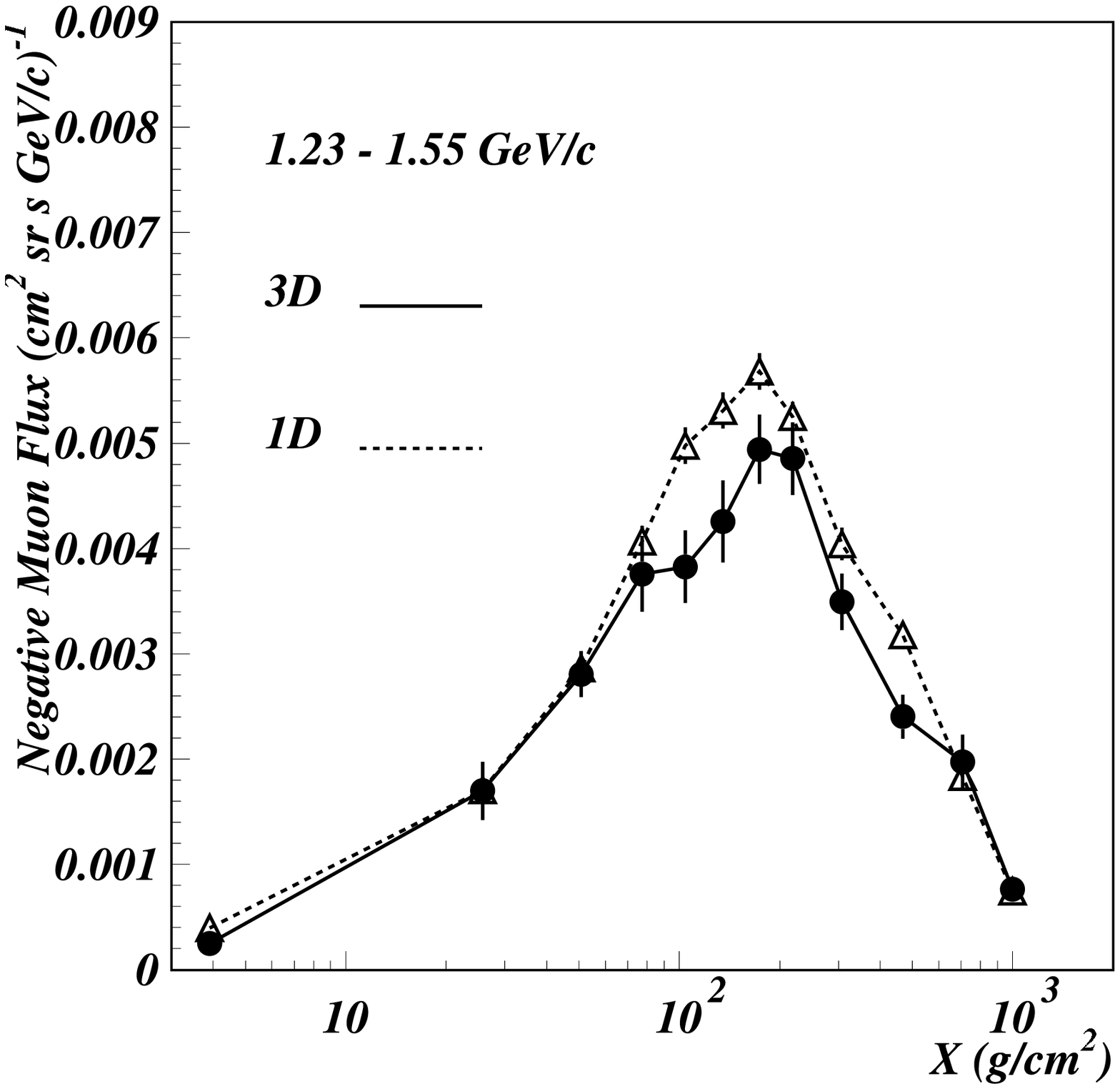,width=7cm}} \\
\end{tabular}
\caption{Comparison between 3--Dimensional and 1--Dimensional
negative muon flux calculations as a function of atmospheric depth
in a few different momentum ranges.
The lines are drawn to guide the eye.
case.\label{fig:1d3d}} 
\end{center}
\end{figure}

\begin{figure}[p]
\begin{center}
\begin{tabular}{cc}
\mbox{\epsfig{file=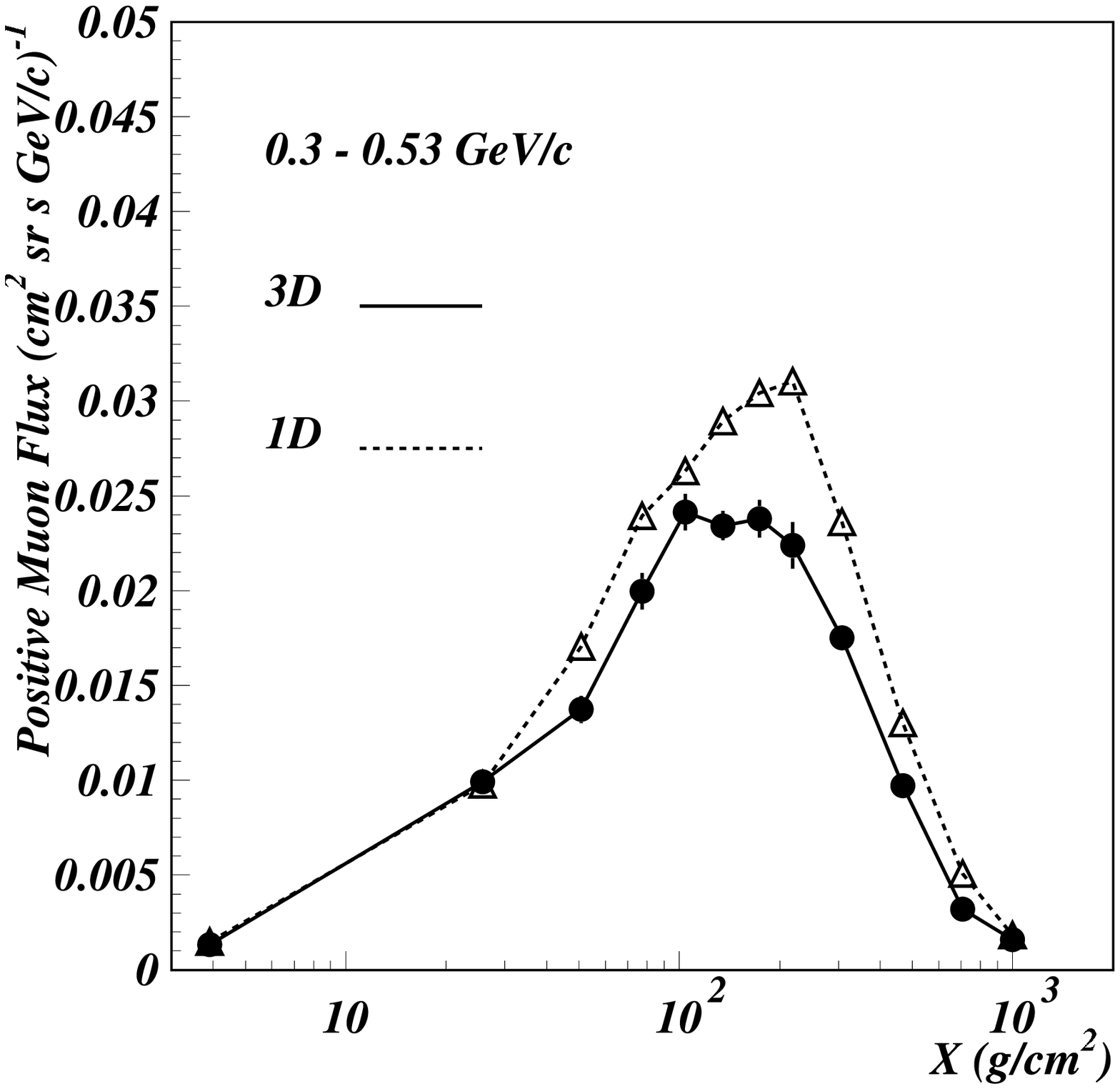,width=7cm}} &
\mbox{\epsfig{file=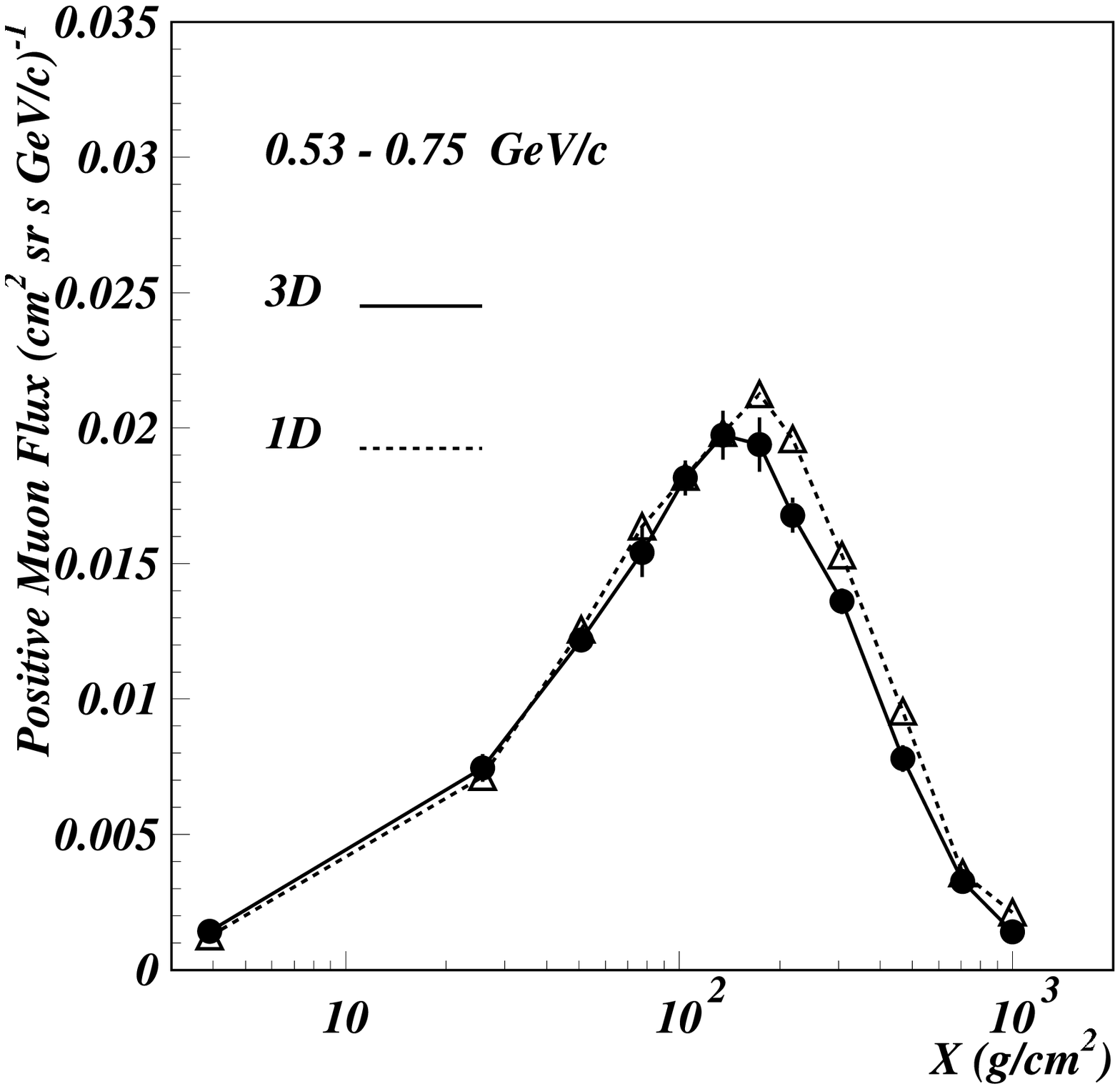,width=7cm}} \\
\mbox{\epsfig{file=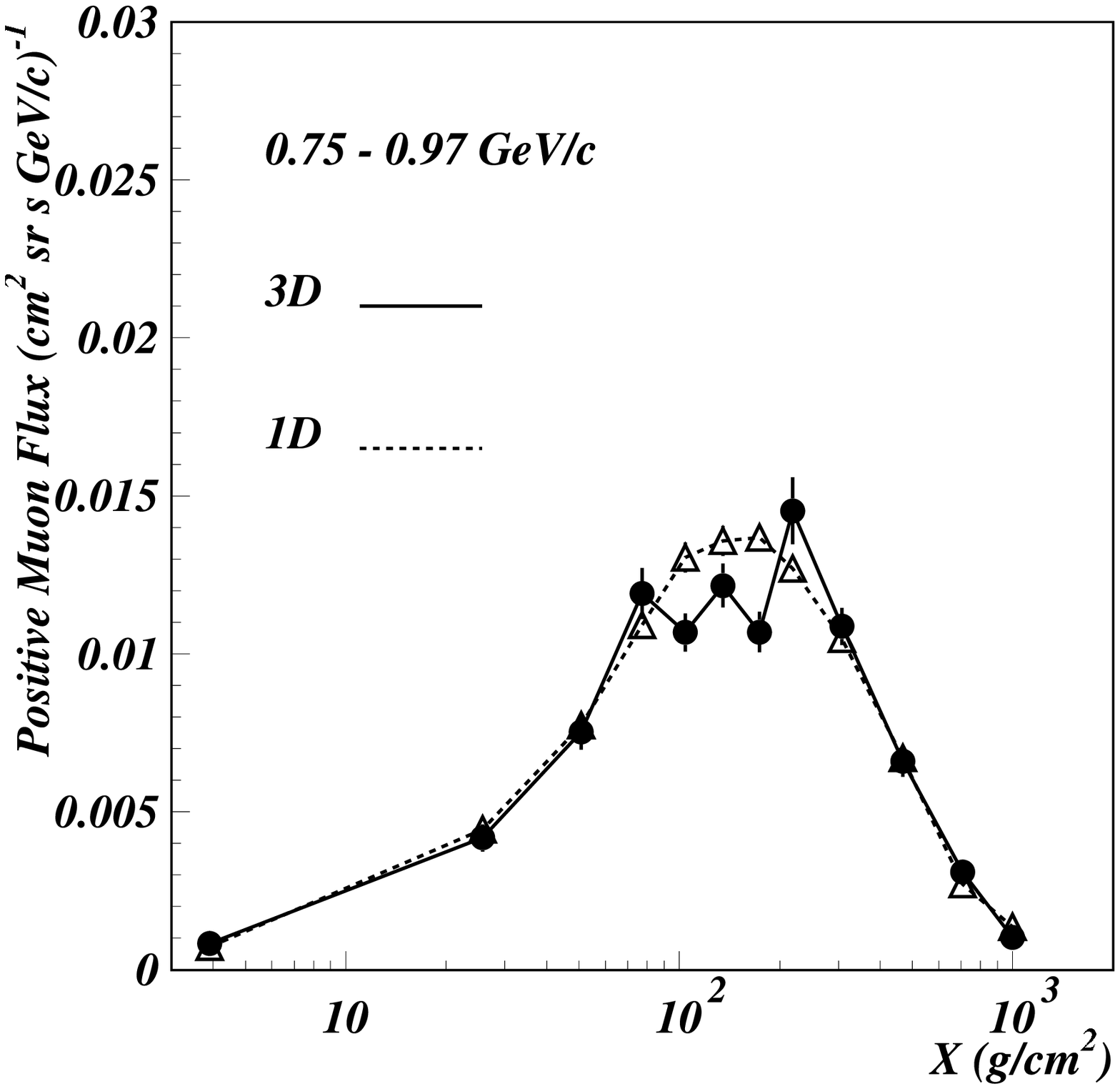,width=7cm}} &
\mbox{\epsfig{file=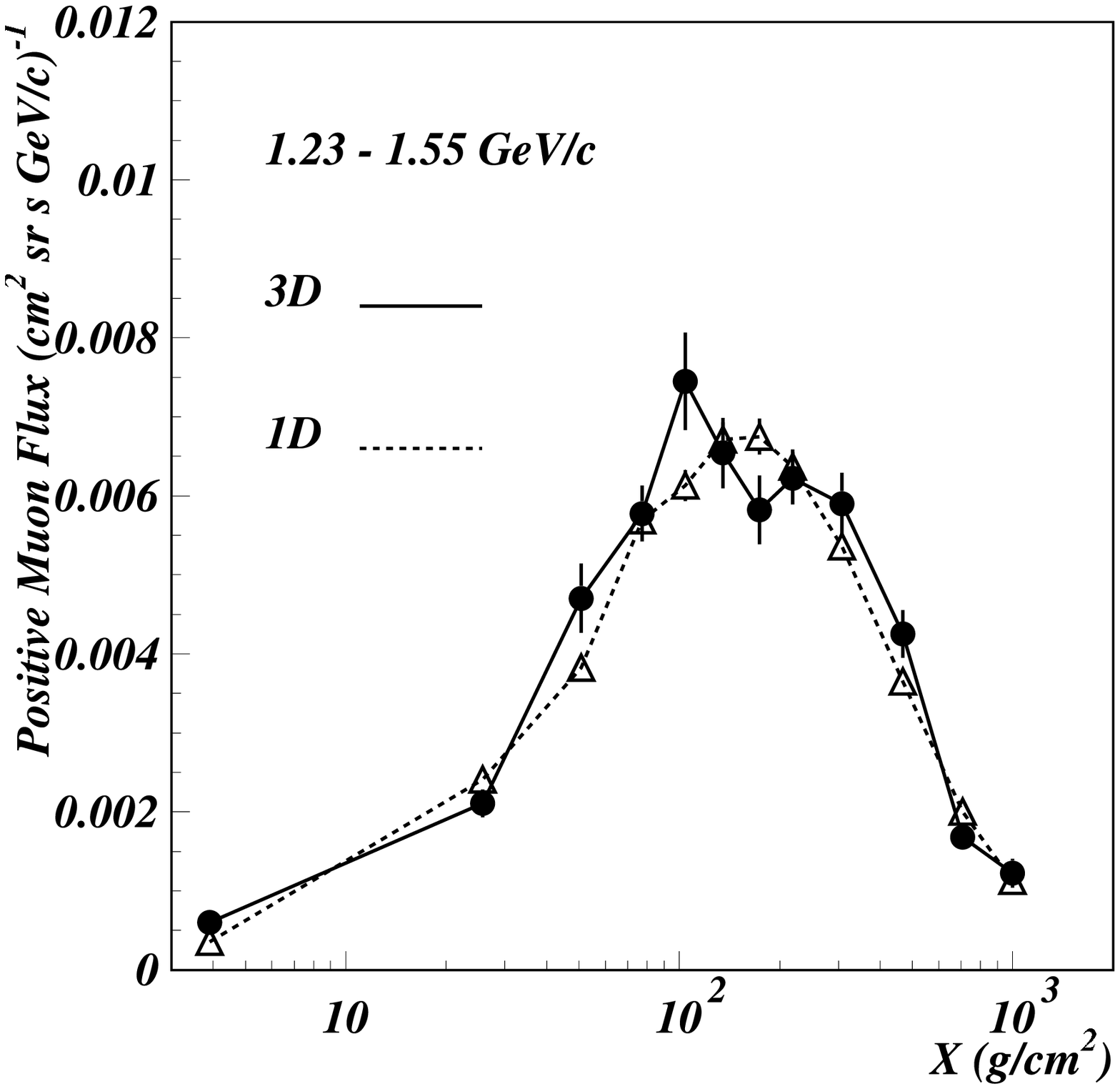,width=7cm}} \\
\end{tabular}
\caption{ Comparison between 3--Dimensional and 1--Dimensional
positive muon flux calculations as a function of atmospheric depth
in a few different momentum ranges.
The lines are drawn to guide the eye.\label{fig:1d3dp}} 
\end{center}
\end{figure}

In order to understand the possible relevance of the 3--dimensional 
simulation setup we have also analyzed data obtained
with the 1--dimensional
runs. From this comparison we can conclude that differences exist at low
muon momenta, where a noticeable excess of muons is produced
in the 1--Dimensional simulations.
As an example, we show in Fig. \ref{fig:1d3d} the
calculated negative muon flux
as a function of atmospheric depth in 4 different momentum intervals
for the 1--Dimensional and 3--Dimensional
simulations. Fig. \ref{fig:1d3dp} shows similar
results for positive muons. In the lowest momentum bins, the
1--D simulation yields a larger muon flux.
We attribute the difference between the 1--D and 3--D cases to
a different path length distribution of muons:
the inclusion of kinematic production
angle, together with particle bending in the geomagnetic field,
brings to an increase of path length and, as a consequence, to a larger
decay probability. 
This argument has been discussed by the Bartol group and other authors in
ref. \cite{stanev}, after that preliminary 
comparisons of the Bartol 1--D results with CAPRICE data were showing
a significant excess in the calculated flux in the low momentum region, as also
reported in \cite{venezia99}. However, in our opinion it is more probable
that the main reason for the excess in that calculation was an excess
of pion multiplicity. In fact the Bartol group is preparing an improved
version of their interaction model which seems to yield charged pion
multiplicities closer to those from the FLUKA model\cite{bartol_icrc1}.
Other differences in Monte Carlo calculations of muons in the
atmosphere between  
the 1--D and 3--D case have been already reported in ref.\cite{coutu},
where the original interaction model from Bartol was used: however, in this
case, 
significant excess 
of calculated muons as compared to HEAT data was observed at high altitude
for a wide range of momentum. This reinforces our interpretation on
the excess of multiplicity in the first Bartol model.

\section{Conclusions}
The results presented here demonstrate that the FLUKA model can reasonably
 reproduce  the muon fluxes in the atmosphere as measured by the CAPRICE
 experiment. 
Of course, the accuracy 
 and reliability of such a comparison depends also on the choice of
 the atmospheric model 
and, mostly, on the assumptions concerning the primary spectrum and the
geomagnetic cutoff.
In particular we are aware that the Bartol all nucleon flux
should be slightly modified in order to be compatible with the
latest data on primaries\cite{bartol_icrc}.
Furthermore, one should also consider 
 that also the present data from balloon experiments might be
 affected by systematic errors which could be $\aprle$ 10\%\cite{circella}.
In this work we are
 are relying on the choices adopted for our calculation of neutrino fluxes
in atmosphere, and therefore we can use the results of the present study to
at least partially validate the neutrino results, due to the strong link
 between muon and neutrino production. 
As a result we can conclude that, having used the same primary
 spectrum, the normalization difference in neutrino flux
 with respect to the Bartol results cannot be attributed to an insufficient
particle production yield in the shower development in the FLUKA model.
As far as absolute normalization is concerned, the muon data considered
 here allow to conclude that the FLUKA calculation should have at maximum
a 10\% error in the neutrino energy range up to few GeVs.
Of course this last statement is strongly
related to the choice of the primary spectrum, and it is not
 possible at this stage to disentangle this factor from the feature of
the FLUKA interaction model, which is otherwise independently benchmarked
 against particle production data.
We wish to stress  that data taken in controlled conditions at accelerators
should remain the main constraint
for any interaction
 model to be used for particle production in atmosphere. 
In particular we remind that 
hadron--Nucleus and Nucleus--Nucleus interactions in  
the energy range 1$\div$30 GeV are still insufficiently known to allow
the construction of a fully reliable model.
The check with
measured
fluxes of different particles in atmosphere remains however
relevant in order to verify that all other ingredients of the simulation
 setup are  taken into account correctly.

Another conclusion of this work is that 3--Dimensional effects
are of some important also for low energy muons.
If we assume that our 3--D vs. 1--D effect is correct, there could be
implications for the results recently presented by Fiorentini et al.,
in \cite{naumov}. However, the significant differences in SubGev neutrino
fluxes between the results of \cite{naumov} and those from the FLUKA
calculation\cite{flukanu} seem hard to explain on the basis of a geometrical
effect or of differences in the primary spectrum. The main difference
should be the interaction model, but then it is still difficult to explain
why different neutrino fluxes arise from calculations yielding reasonable
reproduction of the same data on muon fluxes, a part from technical errors
or, maybe, from important differences in muon energy loss and decay, but
at this stage this is just a guess. As far as FLUKA is
concerned the energy loss models have been obviously
widely cross--checked and verified after many years of practice in 
precision problems like radiation shielding, dosimetry and other 
radio--biological applications. Muon decay with correct polarization effects
has been checked together with the Bartol group.

%Since their work is based on a 1--Dimensional model, according to the
%results reported here, there should be an overestimation of low momentum
%muons. If this is true, the
%agreement of the Fiorentini et al. calculation with muon data at low
%momenta cannot be considered a full validation for neutrino flux
%calculation: a 3--D computation in the same conditions would 
%produce a lower muon flux. As a consequence, there is the possibility
%that the neutrino flux calculated in \cite{naumov} could be too
%low in the sub--GeV, energy region. 

\section*{Acknowledgments}
We wish to thank M. Boezio and M. Circella for the help received 
in understanding the muon experiments and their
data.

\clearpage

\clearpage

\section*{Appendix: Tables of numerical values}

\tablefirsthead{
\hline 
\small
Muon Mom. & \small Flux  & \small Flux &
\small Flux  \\
\small (GeV/c) & \small (cm$^2$ s sr GeV/c)$^-1$ & \small
(m$^2$ s sr GeV/c)$^-1$ & \small (m$^2$ s sr GeV/c)$^-1$  \\
}

\tablehead{\hline 

  \multicolumn{4}{l}{\small\sl continued from previous page}

\\ \hline }

\tabletail{\hline
  \multicolumn{4}{r}{\small\sl continued on next page}
\\\hline}

\tablelasttail{\hline
  \multicolumn{4}{c}{ }
\\ }

\topcaption{Simulated negative muon flux as a function of momentum for the
different depths in atmosphere.\label{tab1}}

\begin{center}
\begin{supertabular}{|c|c|c|c|}
\hline
\multicolumn{1}{|c|}{ } & \multicolumn{1}{|c|}{3.9 g/cm$^2$} &
\multicolumn{1}{|c|}{25.7 g/cm$^2$} & \multicolumn{1}{|c|}{50.7 g/cm$^2$}
\\
\hline
   0.22 & $9.31\cdot 10^{ -4}\pm 36$ \%  & $2.63\cdot 10^{ -3}\pm 19$ \%  & $7.93\cdot 10^{ -3}\pm 14$ \% \\
   0.28 & $5.18\cdot 10^{ -4}\pm 31$ \%  & $6.04\cdot 10^{ -3}\pm 24$ \%  & $6.94\cdot 10^{ -3}\pm 15$ \% \\
   0.36 & $1.92\cdot 10^{ -3}\pm 25$ \%  & $4.55\cdot 10^{ -3}\pm 24$ \%  & $9.45\cdot 10^{ -3}\pm 10$ \% \\
   0.45 & $8.63\cdot 10^{ -4}\pm 25$ \%  & $2.85\cdot 10^{ -3}\pm 18$ \%  & $8.75\cdot 10^{ -3}\pm 12$ \% \\
   0.56 & $1.17\cdot 10^{ -3}\pm 17$ \%  & $2.14\cdot 10^{ -3}\pm 17$ \%  & $4.68\cdot 10^{ -3}\pm 12$ \% \\
   0.71 & $6.33\cdot 10^{ -4}\pm 22$ \%  & $1.85\cdot 10^{ -3}\pm 12$ \%  & $5.15\cdot 10^{ -3}\pm 13$ \% \\
   0.89 & $8.43\cdot 10^{ -4}\pm 35$ \%  & $1.68\cdot 10^{ -3}\pm 12$ \%  & $3.23\cdot 10^{ -3}\pm 12$ \% \\
   1.12 & $4.47\cdot 10^{ -4}\pm 32$ \%  & $9.58\cdot 10^{ -4}\pm 20$ \%  & $2.26\cdot 10^{ -3}\pm 13$ \% \\
   1.42 & $2.52\cdot 10^{ -4}\pm 18$ \%  & $6.83\cdot 10^{ -4}\pm 18$ \%  & $1.70\cdot 10^{ -3}\pm 16$ \% \\
   1.78 & $1.94\cdot 10^{ -4}\pm 24$ \%  & $3.84\cdot 10^{ -4}\pm 19$ \%  & $1.27\cdot 10^{ -3}\pm  9$ \% \\
   2.24 & $1.10\cdot 10^{ -4}\pm 21$ \%  & $1.34\cdot 10^{ -4}\pm 23$ \%  & $6.74\cdot 10^{ -4}\pm 12$ \% \\
   2.83 & $7.03\cdot 10^{ -5}\pm 37$ \%  & $8.21\cdot 10^{ -5}\pm 21$ \%  & $2.46\cdot 10^{ -4}\pm 16$ \% \\
   3.56 & $2.72\cdot 10^{ -5}\pm 24$ \%  & $9.20\cdot 10^{ -5}\pm 22$ \%  & $1.84\cdot 10^{ -4}\pm 15$ \% \\
   4.48 & $2.18\cdot 10^{ -5}\pm 30$ \%  & $5.22\cdot 10^{ -5}\pm 25$ \%  & $1.55\cdot 10^{ -4}\pm 15$ \% \\
   5.64 & $6.62\cdot 10^{ -6}\pm 37$ \%  & $2.66\cdot 10^{ -5}\pm 18$ \%  & $6.18\cdot 10^{ -5}\pm 16$ \% \\
   7.10 & $5.66\cdot 10^{ -6}\pm 26$ \%  & $1.29\cdot 10^{ -5}\pm 23$ \%  & $5.04\cdot 10^{ -5}\pm 23$ \% \\
   8.93 & $1.13\cdot 10^{ -5}\pm 40$ \%  & $1.52\cdot 10^{ -5}\pm 30$ \%  & $3.44\cdot 10^{ -5}\pm 15$ \% \\
  11.25 & $1.35\cdot 10^{ -6}\pm 40$ \%  & $3.83\cdot 10^{ -6}\pm 29$ \%  & $1.00\cdot 10^{ -5}\pm 22$ \% \\
  14.16 & $3.46\cdot 10^{ -7}\pm 64$ \%  & $2.11\cdot 10^{ -6}\pm 30$ \%  & $5.93\cdot 10^{ -6}\pm 19$ \% \\
  17.83 & $3.90\cdot 10^{ -7}\pm 37$ \%  & $1.34\cdot 10^{ -6}\pm 23$ \%  & $2.65\cdot 10^{ -6}\pm 17$ \% \\
  22.44 & $1.37\cdot 10^{ -7}\pm 60$ \%  & $6.22\cdot 10^{ -7}\pm 26$ \%  & $3.11\cdot 10^{ -6}\pm 24$ \% \\
  28.25 & $5.71\cdot 10^{ -8}\pm 63$ \%  & $4.84\cdot 10^{ -7}\pm 22$ \%  & $1.08\cdot 10^{ -6}\pm 30$ \% \\
  35.57 & $7.24\cdot 10^{ -7}\pm 67$ \%  & $8.17\cdot 10^{ -7}\pm 59$ \%  & $1.26\cdot 10^{ -6}\pm 39$ \% \\
  44.77 & $1.12\cdot 10^{ -8}\pm 54$ \%  & $1.56\cdot 10^{ -7}\pm 32$ \%  & $2.41\cdot 10^{ -7}\pm 24$ \% \\
  56.37 & $3.62\cdot 10^{ -8}\pm 49$ \%  & $6.07\cdot 10^{ -8}\pm 46$ \%  & $9.43\cdot 10^{ -8}\pm 38$ \% \\
  70.96 & $5.63\cdot 10^{ -9}\pm 52$ \%  & $2.97\cdot 10^{ -8}\pm 37$ \%  & $1.06\cdot 10^{ -7}\pm 31$ \% \\
  89.34 & $5.20\cdot 10^{-10}\pm 65$ \%  & $1.14\cdot 10^{ -9}\pm 35$ \%  & $8.23\cdot 10^{ -9}\pm 55$ \% \\
 112.47 & $3.95\cdot 10^{-10}\pm 68$ \%  & $2.47\cdot 10^{ -8}\pm 66$ \%  & $3.74\cdot 10^{ -8}\pm 47$ \% \\
 141.59 & $5.13\cdot 10^{-11}\pm 68$ \%  & $1.52\cdot 10^{ -8}\pm 61$ \%  & $1.56\cdot 10^{ -8}\pm 60$ \% \\
 178.25 & $2.60\cdot 10^{-10}\pm 61$ \%  & $1.15\cdot 10^{ -9}\pm 69$ \%  & $5.43\cdot 10^{ -9}\pm 41$ \% \\
\hline

\hline
\multicolumn{1}{|c|}{ } & \multicolumn{1}{|c|}{77.3 g/cm$^2$} &
\multicolumn{1}{|c|}{104.1 g/cm$^2$} & \multicolumn{1}{|c|}{135.3 g/cm$^2$}
\\
\hline
   0.22 & $9.24\cdot 10^{ -3}\pm 12$ \%  & $1.46\cdot 10^{ -2}\pm  7$ \%  & $1.79\cdot 10^{ -2}\pm 12$ \% \\
   0.28 & $1.44\cdot 10^{ -2}\pm 10$ \%  & $1.64\cdot 10^{ -2}\pm 10$ \%  & $1.72\cdot 10^{ -2}\pm 10$ \% \\
   0.36 & $9.90\cdot 10^{ -3}\pm 10$ \%  & $1.78\cdot 10^{ -2}\pm  9$ \%  & $2.14\cdot 10^{ -2}\pm  6$ \% \\
   0.45 & $1.16\cdot 10^{ -2}\pm  9$ \%  & $1.71\cdot 10^{ -2}\pm 10$ \%  & $1.56\cdot 10^{ -2}\pm  9$ \% \\
   0.56 & $9.55\cdot 10^{ -3}\pm  8$ \%  & $1.29\cdot 10^{ -2}\pm  8$ \%  & $1.23\cdot 10^{ -2}\pm  6$ \% \\
   0.71 & $7.40\cdot 10^{ -3}\pm  9$ \%  & $1.06\cdot 10^{ -2}\pm  8$ \%  & $9.94\cdot 10^{ -3}\pm  7$ \% \\
   0.89 & $6.24\cdot 10^{ -3}\pm  9$ \%  & $8.84\cdot 10^{ -3}\pm  8$ \%  & $8.47\cdot 10^{ -3}\pm  7$ \% \\
   1.12 & $3.67\cdot 10^{ -3}\pm  7$ \%  & $6.51\cdot 10^{ -3}\pm  7$ \%  & $6.80\cdot 10^{ -3}\pm  6$ \% \\
   1.42 & $2.80\cdot 10^{ -3}\pm  8$ \%  & $3.76\cdot 10^{ -3}\pm 10$ \%  & $3.83\cdot 10^{ -3}\pm  9$ \% \\
   1.78 & $1.81\cdot 10^{ -3}\pm  9$ \%  & $2.54\cdot 10^{ -3}\pm 10$ \%  & $2.70\cdot 10^{ -3}\pm  7$ \% \\
   2.24 & $9.60\cdot 10^{ -4}\pm 11$ \%  & $1.23\cdot 10^{ -3}\pm 10$ \%  & $1.50\cdot 10^{ -3}\pm  8$ \% \\
   2.83 & $5.56\cdot 10^{ -4}\pm 15$ \%  & $9.43\cdot 10^{ -4}\pm 12$ \%  & $1.18\cdot 10^{ -3}\pm 11$ \% \\
   3.56 & $2.99\cdot 10^{ -4}\pm 14$ \%  & $4.69\cdot 10^{ -4}\pm 10$ \%  & $5.32\cdot 10^{ -4}\pm  8$ \% \\
   4.48 & $3.35\cdot 10^{ -4}\pm 12$ \%  & $4.66\cdot 10^{ -4}\pm 10$ \%  & $5.35\cdot 10^{ -4}\pm  8$ \% \\
   5.64 & $1.16\cdot 10^{ -4}\pm 14$ \%  & $1.64\cdot 10^{ -4}\pm 10$ \%  & $2.27\cdot 10^{ -4}\pm  8$ \% \\
   7.10 & $6.80\cdot 10^{ -5}\pm 18$ \%  & $1.12\cdot 10^{ -4}\pm 14$ \%  & $1.39\cdot 10^{ -4}\pm 11$ \% \\
   8.93 & $5.13\cdot 10^{ -5}\pm 16$ \%  & $6.49\cdot 10^{ -5}\pm  9$ \%  & $7.27\cdot 10^{ -5}\pm  8$ \% \\
  11.25 & $2.55\cdot 10^{ -5}\pm 20$ \%  & $3.41\cdot 10^{ -5}\pm 17$ \%  & $3.66\cdot 10^{ -5}\pm 16$ \% \\
  14.16 & $1.06\cdot 10^{ -5}\pm 15$ \%  & $1.53\cdot 10^{ -5}\pm 14$ \%  & $1.79\cdot 10^{ -5}\pm 13$ \% \\
  17.83 & $4.52\cdot 10^{ -6}\pm 18$ \%  & $5.92\cdot 10^{ -6}\pm 16$ \%  & $7.16\cdot 10^{ -6}\pm 14$ \% \\
  22.44 & $3.39\cdot 10^{ -6}\pm 18$ \%  & $3.86\cdot 10^{ -6}\pm 17$ \%  & $4.81\cdot 10^{ -6}\pm 17$ \% \\
  28.25 & $2.29\cdot 10^{ -6}\pm 21$ \%  & $3.49\cdot 10^{ -6}\pm 20$ \%  & $3.86\cdot 10^{ -6}\pm 19$ \% \\
  35.57 & $2.06\cdot 10^{ -6}\pm 25$ \%  & $2.42\cdot 10^{ -6}\pm 21$ \%  & $2.68\cdot 10^{ -6}\pm 21$ \% \\
  44.77 & $3.88\cdot 10^{ -7}\pm 17$ \%  & $5.05\cdot 10^{ -7}\pm 17$ \%  & $5.54\cdot 10^{ -7}\pm 16$ \% \\
  56.37 & $2.25\cdot 10^{ -7}\pm 31$ \%  & $3.93\cdot 10^{ -7}\pm 25$ \%  & $4.71\cdot 10^{ -7}\pm 22$ \% \\
  70.96 & $1.96\cdot 10^{ -7}\pm 24$ \%  & $2.39\cdot 10^{ -7}\pm 21$ \%  & $2.43\cdot 10^{ -7}\pm 21$ \% \\
  89.34 & $3.37\cdot 10^{ -8}\pm 29$ \%  & $4.87\cdot 10^{ -8}\pm 24$ \%  & $6.65\cdot 10^{ -8}\pm 22$ \% \\
 112.47 & $3.98\cdot 10^{ -8}\pm 44$ \%  & $4.48\cdot 10^{ -8}\pm 39$ \%  & $6.10\cdot 10^{ -8}\pm 32$ \% \\
 141.59 & $2.02\cdot 10^{ -8}\pm 51$ \%  & $3.00\cdot 10^{ -8}\pm 36$ \%  & $3.53\cdot 10^{ -8}\pm 32$ \% \\
 178.25 & $6.61\cdot 10^{ -9}\pm 34$ \%  & $7.53\cdot 10^{ -9}\pm 32$ \%  & $9.32\cdot 10^{ -9}\pm 26$ \% \\
\hline

\hline
\multicolumn{1}{|c|}{ } & \multicolumn{1}{|c|}{173.8 g/cm$^2$} &
\multicolumn{1}{|c|}{218.8 g/cm$^2$} & \multicolumn{1}{|c|}{308.9 g/cm$^2$}
\\
\hline
   0.22 & $1.67\cdot 10^{ -2}\pm 12$ \%  & $1.53\cdot 10^{ -2}\pm 11$ \%  & $1.88\cdot 10^{ -2}\pm 12$ \% \\
   0.28 & $1.80\cdot 10^{ -2}\pm  8$ \%  & $2.16\cdot 10^{ -2}\pm  8$ \%  & $2.12\cdot 10^{ -2}\pm  6$ \% \\
   0.36 & $1.93\cdot 10^{ -2}\pm  8$ \%  & $2.16\cdot 10^{ -2}\pm  6$ \%  & $2.10\cdot 10^{ -2}\pm  9$ \% \\
   0.45 & $1.85\cdot 10^{ -2}\pm  7$ \%  & $1.82\cdot 10^{ -2}\pm  8$ \%  & $1.91\cdot 10^{ -2}\pm  8$ \% \\
   0.56 & $1.41\cdot 10^{ -2}\pm  9$ \%  & $1.64\cdot 10^{ -2}\pm  7$ \%  & $1.48\cdot 10^{ -2}\pm  6$ \% \\
   0.71 & $1.22\cdot 10^{ -2}\pm  6$ \%  & $1.34\cdot 10^{ -2}\pm  7$ \%  & $1.38\cdot 10^{ -2}\pm  6$ \% \\
   0.89 & $9.97\cdot 10^{ -3}\pm  6$ \%  & $1.03\cdot 10^{ -2}\pm  7$ \%  & $9.03\cdot 10^{ -3}\pm  7$ \% \\
   1.12 & $7.34\cdot 10^{ -3}\pm  6$ \%  & $6.54\cdot 10^{ -3}\pm  7$ \%  & $6.77\cdot 10^{ -3}\pm  8$ \% \\
   1.42 & $4.26\cdot 10^{ -3}\pm  9$ \%  & $4.94\cdot 10^{ -3}\pm  7$ \%  & $4.68\cdot 10^{ -3}\pm  5$ \% \\
   1.78 & $2.77\cdot 10^{ -3}\pm  8$ \%  & $2.59\cdot 10^{ -3}\pm  9$ \%  & $3.07\cdot 10^{ -3}\pm  8$ \% \\
   2.24 & $1.66\cdot 10^{ -3}\pm  7$ \%  & $1.86\cdot 10^{ -3}\pm  8$ \%  & $1.86\cdot 10^{ -3}\pm  6$ \% \\
   2.83 & $1.27\cdot 10^{ -3}\pm 10$ \%  & $1.27\cdot 10^{ -3}\pm 10$ \%  & $1.36\cdot 10^{ -3}\pm  7$ \% \\
   3.56 & $6.36\cdot 10^{ -4}\pm  8$ \%  & $7.05\cdot 10^{ -4}\pm  9$ \%  & $8.80\cdot 10^{ -4}\pm  8$ \% \\
   4.48 & $6.03\cdot 10^{ -4}\pm  7$ \%  & $5.28\cdot 10^{ -4}\pm  6$ \%  & $5.60\cdot 10^{ -4}\pm  6$ \% \\
   5.64 & $2.58\cdot 10^{ -4}\pm  7$ \%  & $3.07\cdot 10^{ -4}\pm  8$ \%  & $3.21\cdot 10^{ -4}\pm  7$ \% \\
   7.10 & $1.57\cdot 10^{ -4}\pm 11$ \%  & $1.67\cdot 10^{ -4}\pm 10$ \%  & $1.99\cdot 10^{ -4}\pm 10$ \% \\
   8.93 & $8.34\cdot 10^{ -5}\pm  8$ \%  & $8.62\cdot 10^{ -5}\pm  8$ \%  & $9.93\cdot 10^{ -5}\pm  7$ \% \\
  11.25 & $3.91\cdot 10^{ -5}\pm 15$ \%  & $4.72\cdot 10^{ -5}\pm 12$ \%  & $6.14\cdot 10^{ -5}\pm 10$ \% \\
  14.16 & $1.99\cdot 10^{ -5}\pm 12$ \%  & $2.56\cdot 10^{ -5}\pm 10$ \%  & $2.56\cdot 10^{ -5}\pm 11$ \% \\
  17.83 & $1.07\cdot 10^{ -5}\pm 14$ \%  & $1.00\cdot 10^{ -5}\pm 13$ \%  & $1.24\cdot 10^{ -5}\pm 13$ \% \\
  22.44 & $5.32\cdot 10^{ -6}\pm 15$ \%  & $5.95\cdot 10^{ -6}\pm 14$ \%  & $7.37\cdot 10^{ -6}\pm 13$ \% \\
  28.25 & $4.09\cdot 10^{ -6}\pm 18$ \%  & $5.50\cdot 10^{ -6}\pm 14$ \%  & $5.94\cdot 10^{ -6}\pm 13$ \% \\
  35.57 & $3.45\cdot 10^{ -6}\pm 21$ \%  & $3.58\cdot 10^{ -6}\pm 20$ \%  & $3.98\cdot 10^{ -6}\pm 18$ \% \\
  44.77 & $6.78\cdot 10^{ -7}\pm 15$ \%  & $7.14\cdot 10^{ -7}\pm 14$ \%  & $9.08\cdot 10^{ -7}\pm 14$ \% \\
  56.37 & $6.09\cdot 10^{ -7}\pm 19$ \%  & $6.39\cdot 10^{ -7}\pm 19$ \%  & $7.44\cdot 10^{ -7}\pm 17$ \% \\
  70.96 & $2.78\cdot 10^{ -7}\pm 20$ \%  & $3.10\cdot 10^{ -7}\pm 18$ \%  & $3.46\cdot 10^{ -7}\pm 19$ \% \\
  89.34 & $9.49\cdot 10^{ -8}\pm 28$ \%  & $9.89\cdot 10^{ -8}\pm 27$ \%  & $1.08\cdot 10^{ -7}\pm 27$ \% \\
 112.47 & $6.57\cdot 10^{ -8}\pm 29$ \%  & $7.03\cdot 10^{ -8}\pm 27$ \%  & $7.77\cdot 10^{ -8}\pm 27$ \% \\
 141.59 & $3.61\cdot 10^{ -8}\pm 31$ \%  & $4.09\cdot 10^{ -8}\pm 28$ \%  & $4.62\cdot 10^{ -8}\pm 27$ \% \\
 178.25 & $1.10\cdot 10^{ -8}\pm 23$ \%  & $1.31\cdot 10^{ -8}\pm 23$ \%  & $1.45\cdot 10^{ -8}\pm 22$ \% \\
\hline

\hline
\multicolumn{1}{|c|}{ } & \multicolumn{1}{|c|}{463.7 g/cm$^2$} &
\multicolumn{1}{|c|}{709.0 g/cm$^2$} & \multicolumn{1}{|c|}{1000.0 g/cm$^2$}
\\
\hline
   0.22 & $7.10\cdot 10^{ -3}\pm 15$ \%  & $9.52\cdot 10^{ -3}\pm 21$ \%  & $2.32\cdot 10^{ -3}\pm 13$ \% \\
   0.28 & $1.09\cdot 10^{ -2}\pm 14$ \%  & $8.44\cdot 10^{ -3}\pm  9$ \%  & $4.00\cdot 10^{ -3}\pm 23$ \% \\
   0.36 & $7.59\cdot 10^{ -3}\pm 10$ \%  & $9.48\cdot 10^{ -3}\pm 10$ \%  & $4.40\cdot 10^{ -3}\pm 16$ \% \\
   0.45 & $9.05\cdot 10^{ -3}\pm  9$ \%  & $6.95\cdot 10^{ -3}\pm 10$ \%  & $3.54\cdot 10^{ -3}\pm 14$ \% \\
   0.56 & $8.33\cdot 10^{ -3}\pm  6$ \%  & $7.92\cdot 10^{ -3}\pm  7$ \%  & $2.56\cdot 10^{ -3}\pm 13$ \% \\
   0.71 & $5.87\cdot 10^{ -3}\pm  8$ \%  & $6.06\cdot 10^{ -3}\pm  9$ \%  & $2.39\cdot 10^{ -3}\pm 12$ \% \\
   0.89 & $5.25\cdot 10^{ -3}\pm  8$ \%  & $4.59\cdot 10^{ -3}\pm  6$ \%  & $1.91\cdot 10^{ -3}\pm 11$ \% \\
   1.12 & $3.56\cdot 10^{ -3}\pm  7$ \%  & $3.62\cdot 10^{ -3}\pm  7$ \%  & $1.34\cdot 10^{ -3}\pm  7$ \% \\
   1.42 & $2.40\cdot 10^{ -3}\pm  9$ \%  & $2.16\cdot 10^{ -3}\pm 10$ \%  & $1.98\cdot 10^{ -3}\pm 13$ \% \\
   1.78 & $2.16\cdot 10^{ -3}\pm  9$ \%  & $2.11\cdot 10^{ -3}\pm  8$ \%  & $1.26\cdot 10^{ -3}\pm  7$ \% \\
   2.24 & $1.80\cdot 10^{ -3}\pm  8$ \%  & $1.69\cdot 10^{ -3}\pm  8$ \%  & $9.97\cdot 10^{ -4}\pm  9$ \% \\
   2.83 & $9.90\cdot 10^{ -4}\pm  7$ \%  & $1.03\cdot 10^{ -3}\pm  8$ \%  & $6.98\cdot 10^{ -4}\pm  8$ \% \\
   3.56 & $7.67\cdot 10^{ -4}\pm  8$ \%  & $6.86\cdot 10^{ -4}\pm  7$ \%  & $4.77\cdot 10^{ -4}\pm 10$ \% \\
   4.48 & $3.62\cdot 10^{ -4}\pm  8$ \%  & $3.57\cdot 10^{ -4}\pm  8$ \%  & $3.24\cdot 10^{ -4}\pm  9$ \% \\
   5.64 & $2.99\cdot 10^{ -4}\pm  5$ \%  & $3.00\cdot 10^{ -4}\pm  5$ \%  & $2.13\cdot 10^{ -4}\pm 10$ \% \\
   7.10 & $1.39\cdot 10^{ -4}\pm  8$ \%  & $1.31\cdot 10^{ -4}\pm  9$ \%  & $1.10\cdot 10^{ -4}\pm  9$ \% \\
   8.93 & $8.61\cdot 10^{ -5}\pm  8$ \%  & $8.50\cdot 10^{ -5}\pm  8$ \%  & $6.70\cdot 10^{ -5}\pm  9$ \% \\
  11.25 & $4.88\cdot 10^{ -5}\pm  9$ \%  & $4.81\cdot 10^{ -5}\pm  9$ \%  & $4.23\cdot 10^{ -5}\pm 12$ \% \\
  14.16 & $2.72\cdot 10^{ -5}\pm 10$ \%  & $2.71\cdot 10^{ -5}\pm 10$ \%  & $2.57\cdot 10^{ -5}\pm 11$ \% \\
  17.83 & $1.44\cdot 10^{ -5}\pm 12$ \%  & $1.41\cdot 10^{ -5}\pm 12$ \%  & $1.31\cdot 10^{ -5}\pm 13$ \% \\
  22.44 & $1.08\cdot 10^{ -5}\pm  9$ \%  & $1.08\cdot 10^{ -5}\pm  9$ \%  & $1.05\cdot 10^{ -5}\pm 12$ \% \\
  28.25 & $7.34\cdot 10^{ -6}\pm 13$ \%  & $7.37\cdot 10^{ -6}\pm 13$ \%  & $6.53\cdot 10^{ -6}\pm 12$ \% \\
  35.57 & $3.50\cdot 10^{ -6}\pm 13$ \%  & $3.47\cdot 10^{ -6}\pm 13$ \%  & $3.41\cdot 10^{ -6}\pm 13$ \% \\
  44.77 & $1.29\cdot 10^{ -6}\pm 13$ \%  & $1.40\cdot 10^{ -6}\pm 14$ \%  & $1.35\cdot 10^{ -6}\pm 14$ \% \\
  56.37 & $9.40\cdot 10^{ -7}\pm 15$ \%  & $8.98\cdot 10^{ -7}\pm 16$ \%  & $9.15\cdot 10^{ -7}\pm 15$ \% \\
  70.96 & $3.27\cdot 10^{ -7}\pm 18$ \%  & $3.27\cdot 10^{ -7}\pm 18$ \%  & $3.32\cdot 10^{ -7}\pm 18$ \% \\
  89.34 & $1.67\cdot 10^{ -7}\pm 23$ \%  & $1.71\cdot 10^{ -7}\pm 23$ \%  & $1.66\cdot 10^{ -7}\pm 24$ \% \\
 112.47 & $1.02\cdot 10^{ -7}\pm 21$ \%  & $1.03\cdot 10^{ -7}\pm 21$ \%  & $1.00\cdot 10^{ -7}\pm 22$ \% \\
 141.59 & $5.60\cdot 10^{ -8}\pm 22$ \%  & $5.61\cdot 10^{ -8}\pm 22$ \%  & $5.15\cdot 10^{ -8}\pm 23$ \% \\
 178.25 & $1.74\cdot 10^{ -8}\pm 19$ \%  & $1.76\cdot 10^{ -8}\pm 19$ \%  & $1.75\cdot 10^{ -8}\pm 19$ \% \\
\hline
\end{supertabular}
\end{center}

\clearpage

\tablefirsthead{
\hline 
\small
Muon Mom. & \small Flux  & \small Flux &
\small Flux  \\
\small (GeV/c) & \small (cm$^2$ s sr GeV/c)$^-1$ & \small
(m$^2$ s sr GeV/c)$^-1$ & \small (m$^2$ s sr GeV/c)$^-1$  \\
}

\tablehead{\hline 

  \multicolumn{4}{l}{\small\sl continued from previous page}

\\ \hline }

\tabletail{\hline
  \multicolumn{4}{r}{\small\sl continued on next page}
\\\hline}

\tablelasttail{\hline
  \multicolumn{4}{c}{ }
\\ }

\topcaption{Simulated positive muon flux as a function of momentum for the
different depths in atmosphere.\label{tab2}}

\begin{center}
\begin{supertabular}{|c|c|c|c|}
\hline
\multicolumn{1}{|c|}{ } & \multicolumn{1}{|c|}{3.9 g/cm$^2$} &
\multicolumn{1}{|c|}{25.7 g/cm$^2$} & \multicolumn{1}{|c|}{50.7 g/cm$^2$}
\\
\hline
   0.22 & $8.18\cdot 10^{ -4}\pm 50$ \%  & $7.37\cdot 10^{ -3}\pm 18$ \%  & $1.47\cdot 10^{ -2}\pm  9$ \% \\
   0.28 & $7.41\cdot 10^{ -4}\pm 27$ \%  & $1.08\cdot 10^{ -2}\pm 14$ \%  & $1.09\cdot 10^{ -2}\pm 13$ \% \\
   0.36 & $9.79\cdot 10^{ -4}\pm 27$ \%  & $9.51\cdot 10^{ -3}\pm 11$ \%  & $1.34\cdot 10^{ -2}\pm 10$ \% \\
   0.45 & $1.98\cdot 10^{ -3}\pm 32$ \%  & $9.67\cdot 10^{ -3}\pm 12$ \%  & $1.58\cdot 10^{ -2}\pm  7$ \% \\
   0.56 & $6.18\cdot 10^{ -4}\pm 21$ \%  & $7.39\cdot 10^{ -3}\pm  7$ \%  & $1.31\cdot 10^{ -2}\pm  7$ \% \\
   0.71 & $2.05\cdot 10^{ -3}\pm 19$ \%  & $7.52\cdot 10^{ -3}\pm 12$ \%  & $1.15\cdot 10^{ -2}\pm  7$ \% \\
   0.89 & $8.24\cdot 10^{ -4}\pm 18$ \%  & $4.18\cdot 10^{ -3}\pm 10$ \%  & $7.53\cdot 10^{ -3}\pm  7$ \% \\
   1.12 & $9.47\cdot 10^{ -4}\pm 21$ \%  & $2.81\cdot 10^{ -3}\pm  8$ \%  & $5.07\cdot 10^{ -3}\pm  8$ \% \\
   1.42 & $6.02\cdot 10^{ -4}\pm 20$ \%  & $2.11\cdot 10^{ -3}\pm  8$ \%  & $4.70\cdot 10^{ -3}\pm  9$ \% \\
   1.78 & $2.37\cdot 10^{ -4}\pm 28$ \%  & $1.81\cdot 10^{ -3}\pm 10$ \%  & $2.18\cdot 10^{ -3}\pm 10$ \% \\
   2.24 & $2.20\cdot 10^{ -4}\pm 32$ \%  & $8.33\cdot 10^{ -4}\pm 16$ \%  & $1.51\cdot 10^{ -3}\pm 14$ \% \\
   2.83 & $7.05\cdot 10^{ -5}\pm 23$ \%  & $5.32\cdot 10^{ -4}\pm 14$ \%  & $7.43\cdot 10^{ -4}\pm 10$ \% \\
   3.56 & $8.84\cdot 10^{ -5}\pm 42$ \%  & $3.88\cdot 10^{ -4}\pm 12$ \%  & $5.60\cdot 10^{ -4}\pm 12$ \% \\
   4.48 & $4.09\cdot 10^{ -5}\pm 35$ \%  & $1.50\cdot 10^{ -4}\pm 13$ \%  & $2.88\cdot 10^{ -4}\pm 12$ \% \\
   5.64 & $8.99\cdot 10^{ -6}\pm 39$ \%  & $8.28\cdot 10^{ -5}\pm 11$ \%  & $1.43\cdot 10^{ -4}\pm 10$ \% \\
   7.10 & $1.18\cdot 10^{ -5}\pm 50$ \%  & $4.28\cdot 10^{ -5}\pm 15$ \%  & $8.14\cdot 10^{ -5}\pm 13$ \% \\
   8.93 & $3.02\cdot 10^{ -6}\pm 35$ \%  & $1.82\cdot 10^{ -5}\pm 21$ \%  & $4.00\cdot 10^{ -5}\pm 18$ \% \\
  11.25 & $2.57\cdot 10^{ -6}\pm 33$ \%  & $1.35\cdot 10^{ -5}\pm 19$ \%  & $1.85\cdot 10^{ -5}\pm 17$ \% \\
  14.16 & $1.34\cdot 10^{ -6}\pm 32$ \%  & $4.31\cdot 10^{ -6}\pm 22$ \%  & $7.12\cdot 10^{ -6}\pm 18$ \% \\
  17.83 & $5.71\cdot 10^{ -8}\pm 37$ \%  & $6.60\cdot 10^{ -6}\pm 31$ \%  & $1.05\cdot 10^{ -5}\pm 20$ \% \\
  22.44 & $5.31\cdot 10^{ -7}\pm 29$ \%  & $3.30\cdot 10^{ -6}\pm 18$ \%  & $4.23\cdot 10^{ -6}\pm 15$ \% \\
  28.25 & $8.30\cdot 10^{ -8}\pm 53$ \%  & $1.32\cdot 10^{ -6}\pm 18$ \%  & $1.88\cdot 10^{ -6}\pm 17$ \% \\
  35.57 & $8.85\cdot 10^{ -9}\pm 37$ \%  & $9.68\cdot 10^{ -7}\pm 25$ \%  & $1.44\cdot 10^{ -6}\pm 19$ \% \\
  44.77 & $1.39\cdot 10^{ -8}\pm 46$ \%  & $3.09\cdot 10^{ -7}\pm 30$ \%  & $5.67\cdot 10^{ -7}\pm 22$ \% \\
  56.37 & $2.90\cdot 10^{ -8}\pm 51$ \%  & $1.14\cdot 10^{ -7}\pm 36$ \%  & $3.83\cdot 10^{ -7}\pm 44$ \% \\
  70.96 & $4.52\cdot 10^{ -8}\pm 57$ \%  & $7.48\cdot 10^{ -8}\pm 36$ \%  & $1.21\cdot 10^{ -7}\pm 26$ \% \\
  89.34 & $1.09\cdot 10^{ -8}\pm 84$ \%  & $1.07\cdot 10^{ -7}\pm 64$ \%  & $1.33\cdot 10^{ -7}\pm 51$ \% \\
 112.47 & $1.77\cdot 10^{ -9}\pm 77$ \%  & $3.24\cdot 10^{ -8}\pm 35$ \%  & $5.60\cdot 10^{ -8}\pm 28$ \% \\
 141.59 & $0.00\cdot 10^{  0}\pm  0$ \%  & $3.21\cdot 10^{ -9}\pm 42$ \%  & $6.85\cdot 10^{ -9}\pm 28$ \% \\
 178.25 & $0.00\cdot 10^{  0}\pm  0$ \%  & $1.99\cdot 10^{ -9}\pm 59$ \%  & $2.43\cdot 10^{ -9}\pm 48$ \% \\
\hline

\hline
\multicolumn{1}{|c|}{ } & \multicolumn{1}{|c|}{77.3 g/cm$^2$} &
\multicolumn{1}{|c|}{104.1 g/cm$^2$} & \multicolumn{1}{|c|}{135.3 g/cm$^2$}
\\
\hline
   0.22 & $1.76\cdot 10^{ -2}\pm 11$ \%  & $1.91\cdot 10^{ -2}\pm  9$ \%  & $1.99\cdot 10^{ -2}\pm 12$ \% \\
   0.28 & $1.99\cdot 10^{ -2}\pm  9$ \%  & $2.23\cdot 10^{ -2}\pm  8$ \%  & $2.46\cdot 10^{ -2}\pm  8$ \% \\
   0.36 & $2.20\cdot 10^{ -2}\pm  6$ \%  & $2.47\cdot 10^{ -2}\pm  8$ \%  & $2.51\cdot 10^{ -2}\pm  8$ \% \\
   0.45 & $1.84\cdot 10^{ -2}\pm  7$ \%  & $2.49\cdot 10^{ -2}\pm  5$ \%  & $2.13\cdot 10^{ -2}\pm  5$ \% \\
   0.56 & $1.71\cdot 10^{ -2}\pm  9$ \%  & $1.76\cdot 10^{ -2}\pm  6$ \%  & $2.24\cdot 10^{ -2}\pm  7$ \% \\
   0.71 & $1.41\cdot 10^{ -2}\pm  6$ \%  & $1.86\cdot 10^{ -2}\pm  6$ \%  & $1.76\cdot 10^{ -2}\pm  7$ \% \\
   0.89 & $1.19\cdot 10^{ -2}\pm  7$ \%  & $1.07\cdot 10^{ -2}\pm  6$ \%  & $1.22\cdot 10^{ -2}\pm  6$ \% \\
   1.12 & $8.09\cdot 10^{ -3}\pm 10$ \%  & $9.85\cdot 10^{ -3}\pm  6$ \%  & $1.06\cdot 10^{ -2}\pm  7$ \% \\
   1.42 & $5.78\cdot 10^{ -3}\pm  6$ \%  & $7.45\cdot 10^{ -3}\pm  8$ \%  & $6.54\cdot 10^{ -3}\pm  7$ \% \\
   1.78 & $3.34\cdot 10^{ -3}\pm  9$ \%  & $4.26\cdot 10^{ -3}\pm  9$ \%  & $4.70\cdot 10^{ -3}\pm  8$ \% \\
   2.24 & $2.17\cdot 10^{ -3}\pm  9$ \%  & $2.62\cdot 10^{ -3}\pm  8$ \%  & $2.80\cdot 10^{ -3}\pm 10$ \% \\
   2.83 & $1.11\cdot 10^{ -3}\pm  7$ \%  & $1.32\cdot 10^{ -3}\pm  8$ \%  & $1.24\cdot 10^{ -3}\pm 10$ \% \\
   3.56 & $6.74\cdot 10^{ -4}\pm  8$ \%  & $9.12\cdot 10^{ -4}\pm  7$ \%  & $1.03\cdot 10^{ -3}\pm  7$ \% \\
   4.48 & $3.92\cdot 10^{ -4}\pm 10$ \%  & $4.13\cdot 10^{ -4}\pm 10$ \%  & $4.70\cdot 10^{ -4}\pm  9$ \% \\
   5.64 & $1.95\cdot 10^{ -4}\pm  9$ \%  & $2.15\cdot 10^{ -4}\pm  9$ \%  & $2.67\cdot 10^{ -4}\pm  8$ \% \\
   7.10 & $1.30\cdot 10^{ -4}\pm 14$ \%  & $1.61\cdot 10^{ -4}\pm 12$ \%  & $1.88\cdot 10^{ -4}\pm 10$ \% \\
   8.93 & $5.77\cdot 10^{ -5}\pm 14$ \%  & $6.88\cdot 10^{ -5}\pm 12$ \%  & $8.23\cdot 10^{ -5}\pm 11$ \% \\
  11.25 & $2.46\cdot 10^{ -5}\pm 15$ \%  & $2.63\cdot 10^{ -5}\pm 12$ \%  & $3.10\cdot 10^{ -5}\pm 11$ \% \\
  14.16 & $2.65\cdot 10^{ -5}\pm 21$ \%  & $3.35\cdot 10^{ -5}\pm 20$ \%  & $3.77\cdot 10^{ -5}\pm 16$ \% \\
  17.83 & $1.33\cdot 10^{ -5}\pm 16$ \%  & $1.35\cdot 10^{ -5}\pm 15$ \%  & $1.67\cdot 10^{ -5}\pm 14$ \% \\
  22.44 & $6.27\cdot 10^{ -6}\pm 19$ \%  & $9.10\cdot 10^{ -6}\pm 14$ \%  & $1.03\cdot 10^{ -5}\pm 13$ \% \\
  28.25 & $3.59\cdot 10^{ -6}\pm 16$ \%  & $3.94\cdot 10^{ -6}\pm 15$ \%  & $4.36\cdot 10^{ -6}\pm 15$ \% \\
  35.57 & $1.81\cdot 10^{ -6}\pm 19$ \%  & $3.24\cdot 10^{ -6}\pm 21$ \%  & $3.27\cdot 10^{ -6}\pm 21$ \% \\
  44.77 & $8.21\cdot 10^{ -7}\pm 21$ \%  & $9.28\cdot 10^{ -7}\pm 20$ \%  & $9.98\cdot 10^{ -7}\pm 18$ \% \\
  56.37 & $4.88\cdot 10^{ -7}\pm 35$ \%  & $9.58\cdot 10^{ -7}\pm 34$ \%  & $1.05\cdot 10^{ -6}\pm 30$ \% \\
  70.96 & $1.43\cdot 10^{ -7}\pm 22$ \%  & $1.91\cdot 10^{ -7}\pm 21$ \%  & $2.19\cdot 10^{ -7}\pm 19$ \% \\
  89.34 & $1.59\cdot 10^{ -7}\pm 43$ \%  & $1.74\cdot 10^{ -7}\pm 39$ \%  & $1.92\cdot 10^{ -7}\pm 35$ \% \\
 112.47 & $6.66\cdot 10^{ -8}\pm 24$ \%  & $6.99\cdot 10^{ -8}\pm 22$ \%  & $7.90\cdot 10^{ -8}\pm 20$ \% \\
 141.59 & $1.28\cdot 10^{ -8}\pm 21$ \%  & $1.38\cdot 10^{ -8}\pm 21$ \%  & $1.74\cdot 10^{ -8}\pm 17$ \% \\
 178.25 & $3.13\cdot 10^{ -9}\pm 38$ \%  & $3.63\cdot 10^{ -9}\pm 33$ \%  & $3.93\cdot 10^{ -9}\pm 30$ \% \\
\hline

\hline
\multicolumn{1}{|c|}{ } & \multicolumn{1}{|c|}{173.8 g/cm$^2$} &
\multicolumn{1}{|c|}{218.8 g/cm$^2$} & \multicolumn{1}{|c|}{308.9 g/cm$^2$}
\\
\hline
   0.22 & $2.41\cdot 10^{ -2}\pm 11$ \%  & $2.01\cdot 10^{ -2}\pm  9$ \%  & $1.63\cdot 10^{ -2}\pm  7$ \% \\
   0.28 & $2.32\cdot 10^{ -2}\pm  8$ \%  & $2.28\cdot 10^{ -2}\pm  7$ \%  & $1.56\cdot 10^{ -2}\pm 10$ \% \\
   0.36 & $2.76\cdot 10^{ -2}\pm  5$ \%  & $2.03\cdot 10^{ -2}\pm 10$ \%  & $1.55\cdot 10^{ -2}\pm  7$ \% \\
   0.45 & $2.12\cdot 10^{ -2}\pm  7$ \%  & $2.38\cdot 10^{ -2}\pm  6$ \%  & $2.03\cdot 10^{ -2}\pm  8$ \% \\
   0.56 & $2.30\cdot 10^{ -2}\pm  8$ \%  & $1.95\cdot 10^{ -2}\pm  5$ \%  & $1.38\cdot 10^{ -2}\pm  4$ \% \\
   0.71 & $1.65\cdot 10^{ -2}\pm  5$ \%  & $1.46\cdot 10^{ -2}\pm  6$ \%  & $1.35\cdot 10^{ -2}\pm  6$ \% \\
   0.89 & $1.07\cdot 10^{ -2}\pm  6$ \%  & $1.45\cdot 10^{ -2}\pm  7$ \%  & $1.09\cdot 10^{ -2}\pm  5$ \% \\
   1.12 & $1.14\cdot 10^{ -2}\pm  5$ \%  & $9.80\cdot 10^{ -3}\pm  5$ \%  & $7.66\cdot 10^{ -3}\pm  7$ \% \\
   1.42 & $5.82\cdot 10^{ -3}\pm  8$ \%  & $6.22\cdot 10^{ -3}\pm  5$ \%  & $5.90\cdot 10^{ -3}\pm  7$ \% \\
   1.78 & $4.80\cdot 10^{ -3}\pm  7$ \%  & $4.81\cdot 10^{ -3}\pm  6$ \%  & $4.44\cdot 10^{ -3}\pm  9$ \% \\
   2.24 & $2.68\cdot 10^{ -3}\pm  7$ \%  & $2.36\cdot 10^{ -3}\pm  5$ \%  & $2.08\cdot 10^{ -3}\pm  9$ \% \\
   2.83 & $1.47\cdot 10^{ -3}\pm  9$ \%  & $1.72\cdot 10^{ -3}\pm  7$ \%  & $1.43\cdot 10^{ -3}\pm  7$ \% \\
   3.56 & $1.10\cdot 10^{ -3}\pm  6$ \%  & $9.81\cdot 10^{ -4}\pm  8$ \%  & $9.55\cdot 10^{ -4}\pm  7$ \% \\
   4.48 & $5.58\cdot 10^{ -4}\pm 10$ \%  & $5.23\cdot 10^{ -4}\pm  9$ \%  & $5.03\cdot 10^{ -4}\pm  7$ \% \\
   5.64 & $3.08\cdot 10^{ -4}\pm  7$ \%  & $3.30\cdot 10^{ -4}\pm  7$ \%  & $3.23\cdot 10^{ -4}\pm  9$ \% \\
   7.10 & $1.91\cdot 10^{ -4}\pm  9$ \%  & $2.09\cdot 10^{ -4}\pm  7$ \%  & $2.31\cdot 10^{ -4}\pm  8$ \% \\
   8.93 & $8.20\cdot 10^{ -5}\pm 12$ \%  & $9.27\cdot 10^{ -5}\pm 10$ \%  & $1.07\cdot 10^{ -4}\pm  8$ \% \\
  11.25 & $4.10\cdot 10^{ -5}\pm 10$ \%  & $4.26\cdot 10^{ -5}\pm 10$ \%  & $5.97\cdot 10^{ -5}\pm  7$ \% \\
  14.16 & $4.17\cdot 10^{ -5}\pm 14$ \%  & $4.78\cdot 10^{ -5}\pm 13$ \%  & $4.84\cdot 10^{ -5}\pm 10$ \% \\
  17.83 & $2.35\cdot 10^{ -5}\pm 11$ \%  & $2.61\cdot 10^{ -5}\pm 11$ \%  & $2.97\cdot 10^{ -5}\pm 11$ \% \\
  22.44 & $1.11\cdot 10^{ -5}\pm 12$ \%  & $1.22\cdot 10^{ -5}\pm 11$ \%  & $1.21\cdot 10^{ -5}\pm 10$ \% \\
  28.25 & $4.86\cdot 10^{ -6}\pm 14$ \%  & $5.48\cdot 10^{ -6}\pm 14$ \%  & $5.99\cdot 10^{ -6}\pm 12$ \% \\
  35.57 & $4.24\cdot 10^{ -6}\pm 18$ \%  & $5.17\cdot 10^{ -6}\pm 16$ \%  & $5.36\cdot 10^{ -6}\pm 16$ \% \\
  44.77 & $1.16\cdot 10^{ -6}\pm 17$ \%  & $1.20\cdot 10^{ -6}\pm 16$ \%  & $1.28\cdot 10^{ -6}\pm 15$ \% \\
  56.37 & $1.07\cdot 10^{ -6}\pm 30$ \%  & $1.21\cdot 10^{ -6}\pm 26$ \%  & $1.29\cdot 10^{ -6}\pm 25$ \% \\
  70.96 & $2.49\cdot 10^{ -7}\pm 18$ \%  & $3.87\cdot 10^{ -7}\pm 24$ \%  & $3.99\cdot 10^{ -7}\pm 22$ \% \\
  89.34 & $2.03\cdot 10^{ -7}\pm 33$ \%  & $1.38\cdot 10^{ -7}\pm 16$ \%  & $1.77\cdot 10^{ -7}\pm 17$ \% \\
 112.47 & $8.56\cdot 10^{ -8}\pm 19$ \%  & $1.01\cdot 10^{ -7}\pm 16$ \%  & $1.03\cdot 10^{ -7}\pm 13$ \% \\
 141.59 & $2.02\cdot 10^{ -8}\pm 15$ \%  & $2.48\cdot 10^{ -8}\pm 16$ \%  & $3.41\cdot 10^{ -8}\pm 16$ \% \\
 178.25 & $4.51\cdot 10^{ -9}\pm 26$ \%  & $5.38\cdot 10^{ -9}\pm 23$ \%  & $6.58\cdot 10^{ -9}\pm 21$ \% \\
\hline

\hline
\multicolumn{1}{|c|}{ } & \multicolumn{1}{|c|}{463.7 g/cm$^2$} &
\multicolumn{1}{|c|}{709.0 g/cm$^2$} & \multicolumn{1}{|c|}{1000.0 g/cm$^2$}
\\
\hline
   0.22 & $7.57\cdot 10^{ -3}\pm 12$ \%  & $3.69\cdot 10^{ -3}\pm 26$ \%  & $7.71\cdot 10^{ -4}\pm 23$ \% \\
   0.28 & $9.69\cdot 10^{ -3}\pm 13$ \%  & $2.27\cdot 10^{ -3}\pm 16$ \%  & $1.47\cdot 10^{ -3}\pm 28$ \% \\
   0.36 & $9.22\cdot 10^{ -3}\pm 11$ \%  & $3.61\cdot 10^{ -3}\pm 11$ \%  & $1.56\cdot 10^{ -3}\pm 18$ \% \\
   0.45 & $1.01\cdot 10^{ -2}\pm  9$ \%  & $3.45\cdot 10^{ -3}\pm 17$ \%  & $1.68\cdot 10^{ -3}\pm 21$ \% \\
   0.56 & $7.99\cdot 10^{ -3}\pm  8$ \%  & $3.66\cdot 10^{ -3}\pm 11$ \%  & $1.48\cdot 10^{ -3}\pm 16$ \% \\
   0.71 & $7.65\cdot 10^{ -3}\pm 11$ \%  & $2.94\cdot 10^{ -3}\pm  8$ \%  & $1.33\cdot 10^{ -3}\pm 16$ \% \\
   0.89 & $6.60\cdot 10^{ -3}\pm  7$ \%  & $3.09\cdot 10^{ -3}\pm 12$ \%  & $1.02\cdot 10^{ -3}\pm 11$ \% \\
   1.12 & $4.67\cdot 10^{ -3}\pm  6$ \%  & $2.31\cdot 10^{ -3}\pm  8$ \%  & $1.12\cdot 10^{ -3}\pm 14$ \% \\
   1.42 & $4.25\cdot 10^{ -3}\pm  7$ \%  & $1.68\cdot 10^{ -3}\pm  9$ \%  & $1.23\cdot 10^{ -3}\pm 15$ \% \\
   1.78 & $2.24\cdot 10^{ -3}\pm  5$ \%  & $1.28\cdot 10^{ -3}\pm 11$ \%  & $7.07\cdot 10^{ -4}\pm 12$ \% \\
   2.24 & $1.91\cdot 10^{ -3}\pm  9$ \%  & $1.20\cdot 10^{ -3}\pm  9$ \%  & $5.80\cdot 10^{ -4}\pm 12$ \% \\
   2.83 & $1.28\cdot 10^{ -3}\pm  6$ \%  & $6.84\cdot 10^{ -4}\pm  9$ \%  & $5.15\cdot 10^{ -4}\pm 13$ \% \\
   3.56 & $6.77\cdot 10^{ -4}\pm  7$ \%  & $5.11\cdot 10^{ -4}\pm  9$ \%  & $3.68\cdot 10^{ -4}\pm 10$ \% \\
   4.48 & $4.79\cdot 10^{ -4}\pm  9$ \%  & $3.48\cdot 10^{ -4}\pm 10$ \%  & $3.04\cdot 10^{ -4}\pm  9$ \% \\
   5.64 & $3.22\cdot 10^{ -4}\pm  8$ \%  & $2.57\cdot 10^{ -4}\pm  8$ \%  & $2.05\cdot 10^{ -4}\pm  8$ \% \\
   7.10 & $1.95\cdot 10^{ -4}\pm  7$ \%  & $1.42\cdot 10^{ -4}\pm  9$ \%  & $8.94\cdot 10^{ -5}\pm 11$ \% \\
   8.93 & $8.71\cdot 10^{ -5}\pm  9$ \%  & $7.82\cdot 10^{ -5}\pm 10$ \%  & $6.78\cdot 10^{ -5}\pm  9$ \% \\
  11.25 & $6.17\cdot 10^{ -5}\pm  7$ \%  & $5.93\cdot 10^{ -5}\pm 10$ \%  & $5.92\cdot 10^{ -5}\pm 14$ \% \\
  14.16 & $5.42\cdot 10^{ -5}\pm 12$ \%  & $4.91\cdot 10^{ -5}\pm 12$ \%  & $3.62\cdot 10^{ -5}\pm 10$ \% \\
  17.83 & $2.48\cdot 10^{ -5}\pm 11$ \%  & $2.13\cdot 10^{ -5}\pm 14$ \%  & $2.02\cdot 10^{ -5}\pm 17$ \% \\
  22.44 & $1.21\cdot 10^{ -5}\pm 10$ \%  & $1.13\cdot 10^{ -5}\pm 11$ \%  & $9.70\cdot 10^{ -6}\pm 13$ \% \\
  28.25 & $6.08\cdot 10^{ -6}\pm 11$ \%  & $7.42\cdot 10^{ -6}\pm 13$ \%  & $7.21\cdot 10^{ -6}\pm 14$ \% \\
  35.57 & $5.60\cdot 10^{ -6}\pm 20$ \%  & $4.27\cdot 10^{ -6}\pm 17$ \%  & $3.96\cdot 10^{ -6}\pm 15$ \% \\
  44.77 & $1.14\cdot 10^{ -6}\pm 16$ \%  & $1.15\cdot 10^{ -6}\pm 16$ \%  & $9.98\cdot 10^{ -7}\pm 15$ \% \\
  56.37 & $1.35\cdot 10^{ -6}\pm 24$ \%  & $1.33\cdot 10^{ -6}\pm 24$ \%  & $1.30\cdot 10^{ -6}\pm 25$ \% \\
  70.96 & $4.08\cdot 10^{ -7}\pm 22$ \%  & $4.05\cdot 10^{ -7}\pm 22$ \%  & $4.06\cdot 10^{ -7}\pm 22$ \% \\
  89.34 & $2.19\cdot 10^{ -7}\pm 15$ \%  & $2.70\cdot 10^{ -7}\pm 15$ \%  & $2.61\cdot 10^{ -7}\pm 16$ \% \\
 112.47 & $8.56\cdot 10^{ -8}\pm 14$ \%  & $8.55\cdot 10^{ -8}\pm 15$ \%  & $9.34\cdot 10^{ -8}\pm 15$ \% \\
 141.59 & $3.48\cdot 10^{ -8}\pm 16$ \%  & $3.59\cdot 10^{ -8}\pm 15$ \%  & $2.88\cdot 10^{ -8}\pm 14$ \% \\
 178.25 & $1.55\cdot 10^{ -8}\pm 22$ \%  & $1.56\cdot 10^{ -8}\pm 23$ \%  & $1.57\cdot 10^{ -8}\pm 22$ \% \\
\hline
\end{supertabular}
\end{center}

\clearpage

\tablefirsthead{
\hline 
\small
Muon Mom. & \small $\mu^+$ Flux  & \small $\mu^-$ Flux \\
\small (GeV/c) & \small (cm$^2$ s sr GeV/c)$^-1$  & \small
(cm$^2$ s sr GeV/c)$^-1$ \\
}

\tablehead{\hline 

  \multicolumn{3}{l}{\small\sl continued from previous page}

\\ \hline }

\tabletail{\hline
  \multicolumn{3}{r}{\small\sl continued on next page}
\\\hline}

\tablelasttail{\hline
  \multicolumn{3}{c}{ }
\\ }

\topcaption{Simulated  muon flux as a function of momentum at ground
depth.\label{tab3}}

\begin{center}
\begin{supertabular}{|c|c|c|}
\hline
   0.22 & $7.71\cdot 10^{ -4}\pm 23$ \%  & $7.79\cdot 10^{ -4}\pm 24$ \%  \\
   0.28 & $1.47\cdot 10^{ -3}\pm 28$ \%  & $8.05\cdot 10^{ -4}\pm 17$ \%  \\
   0.36 & $1.56\cdot 10^{ -3}\pm 18$ \%  & $1.09\cdot 10^{ -3}\pm 20$ \%  \\
   0.45 & $1.68\cdot 10^{ -3}\pm 21$ \%  & $1.64\cdot 10^{ -3}\pm 16$ \%  \\
   0.56 & $1.48\cdot 10^{ -3}\pm 16$ \%  & $8.01\cdot 10^{ -4}\pm 15$ \%  \\
   0.71 & $1.33\cdot 10^{ -3}\pm 16$ \%  & $1.38\cdot 10^{ -3}\pm 10$ \%  \\
   0.89 & $1.02\cdot 10^{ -3}\pm 11$ \%  & $1.29\cdot 10^{ -3}\pm 16$ \%  \\
   1.12 & $1.12\cdot 10^{ -3}\pm 14$ \%  & $1.26\cdot 10^{ -3}\pm 14$ \%  \\
   1.42 & $1.23\cdot 10^{ -3}\pm 15$ \%  & $7.66\cdot 10^{ -4}\pm 12$ \%  \\
   1.78 & $7.07\cdot 10^{ -4}\pm 12$ \%  & $7.35\cdot 10^{ -4}\pm 12$ \%  \\
   2.24 & $5.80\cdot 10^{ -4}\pm 12$ \%  & $6.77\cdot 10^{ -4}\pm 11$ \%  \\
   2.83 & $5.15\cdot 10^{ -4}\pm 13$ \%  & $4.32\cdot 10^{ -4}\pm 12$ \%  \\
   3.56 & $3.68\cdot 10^{ -4}\pm 10$ \%  & $3.21\cdot 10^{ -4}\pm  9$ \%  \\
   4.48 & $3.04\cdot 10^{ -4}\pm  9$ \%  & $2.88\cdot 10^{ -4}\pm 10$ \%  \\
   5.64 & $2.05\cdot 10^{ -4}\pm  8$ \%  & $1.31\cdot 10^{ -4}\pm 11$ \%  \\
   7.10 & $8.94\cdot 10^{ -5}\pm 11$ \%  & $9.28\cdot 10^{ -5}\pm  8$ \%  \\
   8.93 & $6.78\cdot 10^{ -5}\pm  9$ \%  & $5.49\cdot 10^{ -5}\pm  9$ \%  \\
  11.25 & $5.92\cdot 10^{ -5}\pm 14$ \%  & $3.68\cdot 10^{ -5}\pm 11$ \%  \\
  14.16 & $3.62\cdot 10^{ -5}\pm 10$ \%  & $1.98\cdot 10^{ -5}\pm 13$ \%  \\
  17.83 & $2.02\cdot 10^{ -5}\pm 17$ \%  & $1.32\cdot 10^{ -5}\pm 12$ \%  \\
  22.44 & $9.70\cdot 10^{ -6}\pm 13$ \%  & $9.17\cdot 10^{ -6}\pm 13$ \%  \\
  28.25 & $7.21\cdot 10^{ -6}\pm 14$ \%  & $5.39\cdot 10^{ -6}\pm 15$ \%  \\
  35.57 & $3.96\cdot 10^{ -6}\pm 15$ \%  & $3.39\cdot 10^{ -6}\pm 14$ \%  \\
  44.77 & $9.98\cdot 10^{ -7}\pm 15$ \%  & $1.27\cdot 10^{ -6}\pm 14$ \%  \\
  56.37 & $1.30\cdot 10^{ -6}\pm 25$ \%  & $8.44\cdot 10^{ -7}\pm 15$ \%  \\
  70.96 & $4.06\cdot 10^{ -7}\pm 22$ \%  & $3.27\cdot 10^{ -7}\pm 18$ \%  \\
  89.34 & $2.61\cdot 10^{ -7}\pm 16$ \%  & $1.69\cdot 10^{ -7}\pm 24$ \%  \\
 112.47 & $9.34\cdot 10^{ -8}\pm 15$ \%  & $9.80\cdot 10^{ -8}\pm 21$ \%  \\
 141.59 & $2.88\cdot 10^{ -8}\pm 14$ \%  & $5.12\cdot 10^{ -8}\pm 23$ \%  \\
 178.25 & $1.57\cdot 10^{ -8}\pm 22$ \%  & $1.49\cdot 10^{ -8}\pm 19$ \%  \\
\hline
\end{supertabular}
\end{center}


\begin{thebibliography}{9}
\bibitem{fluka} A.~Ferrari and P.R.~Sala, ATLAS internal note
ATL-PHYS-97-113 (1997) accessible through the CERN preprint server;
Proc. of the {\it Workshop on Nuclear Reaction Data and Nuclear Reactors
Physics, Design and Safety}, ICTP, 
Miramare-Trieste, Italy, 15~April--17~May~1996, Proceedings
published by World Scientific, A.~Gandini, G.~Reffo eds, Vol.~2, p.~424,
(1998).
\bibitem{flukanu} G. Battistoni et al., Nuclear Phys. B (Proc. Suppl.) {\bf
70} (1998) 358; G.~Battistoni et al., Astrop. Phys. {\bf 12} (2000) 315.
FLUKA~flux~tables~are~available~in
http://www.mi.infn.it/\~battist/neutrino.html.
\bibitem{bartol} V.~Agrawal et al., Phys. Rev. {\bf D53} (1996) 1314;
G. Barr et al., Phys. Rev. {\bf D39} (1989) 3532;
T. K. Gaisser and T. Stanev, 
Proc. of the 24th ICRC (Rome, 1995), vol.{\bf 1}, 694.
\bibitem{now2000} G. Battistoni, hep-ph/0012268, and
Nucl. Phys. Proc. Suppl. {\bf 100} (2001) 101 {\it (Proc of the
Neutrino Oscillation Workshop 2000, Otranto, Sep. 2000)}.
\bibitem{HKKM} M.~Honda et al., Phys. Rev. {\bf D52} (1995) 4985.
\bibitem{caprice94} M. Boezio et al., Phys. Rev. {\bf D62} (2000) 032007.;
J. Kremer et al., Phys. Rev. Lett. {\bf 83} (1999) 4241.
\bibitem{naumov} G.~Fiorentini, V.A.~Naumov, F.L.~Villante, hep-ph/0103322.
\bibitem{bess} T. Sanuki et al., Astrophys. J. {\bf 545} (2000) 1135.
\bibitem{ams} J. Alcaraz et al., Phys. Lett. {\bf B472} (2000) 215.
\bibitem{wiebel} B. Wiebel-Sooth, P. Biermann and H. Meyer, astro-ph/9709253
\bibitem{modulation} G.D.~Badhwar and P.M.~O'Neill, Adv. Space
Res. Vol. {\bf 17}, No. 2 (1996) 7.
\bibitem{geomag} In the FLUKA calculations of particle fluxes in atmosphere
we have used the IGRF (International Geomagnetic Reference Field)
available from NASA/NSSDC.
\bibitem{atmo} The fit to atmospheric profile has been taken from 
T. K. Gaisser, ``Cosmic Rays and Particle Physics'',  Cambridge
University Press, Cambridge, England. 
\bibitem{stanev} T. Stanev, S. Coutu, T.K. Gaisser and G. Barr, Proc. of 
the 26th ICRC, Utah 1999, {\bf Vol. 2}, 96.
\bibitem{venezia99} M. Circella, talk at the VIII Workshop on Neutrino
Telescope, Venezia, Feb. 1999.
\bibitem{bartol_icrc1} R. Engel, T.K. Gaisser, P. Lipari and T. Stanev,  
Proc. of the 27th ICRC (Hamburg, 2001), Session HE2.02.
\bibitem{coutu} S. Coutu et al., Phys. Rev. {\bf D62} (2000) 032001.
\bibitem{circella} M. Circella, private communication.
\bibitem{bartol_icrc} 
T.K. Gaisser, M. Honda, P. Lipari and T. Stanev, Proc. of the 27th ICRC
(Hamburg, 2001), Session OG1.01


\end{thebibliography}
\end{document}